\documentclass[hidelinks,onefignum,onetabnum]{siamart250211}

\usepackage{lipsum}
\usepackage{amsfonts}
\usepackage{graphicx}
\usepackage{hyperref}
\usepackage{cases}
\usepackage{epstopdf}
\usepackage{algorithmic}
\usepackage{enumitem}
\ifpdf
  \DeclareGraphicsExtensions{.eps,.pdf,.png,.jpg}
\else
  \DeclareGraphicsExtensions{.eps}
\fi
\usepackage{lineno}
\newsiamremark{remark}{Remark}
\newsiamremark{hypothesis}{Hypothesis}
\crefname{hypothesis}{Hypothesis}{Hypotheses}
\newsiamthm{claim}{Claim}
\newsiamremark{fact}{Fact}
\crefname{fact}{Fact}{Facts}

\headers{Coherent dynamics in soft-threshold integrate-and-fire networks}{L. Forbes, J. Grossman, M. Avery, R. Goh, and G.K. Ocker}

\title{Coherent Dynamics in Networks of Soft-threshold Integrate-and-Fire Neurons on the Ring %\thanks{Submitted to the editors \textcolor{red}{DATE}.
}

\author{Lauren Forbes\thanks{Boston University, Department of Mathematics and Statistics
  (\email{lcforbes@bu.edu}, \email{rgoh@bu.edu})}
\and Jared Grossman\footnotemark[1]
\and Montie Avery\footnotemark[1]
\and Ryan Goh\footnotemark[1] 
\and Gabriel Koch Ocker\footnotemark[1] \thanks{and Center for Systems Neuroscience (\email{gkocker@bu.edu}).} %\footnotemark[2]
}

\usepackage{amsopn}

\ifpdf
\hypersetup{
  pdftitle={Coherent Dynamics in Networks of sLIF Neurons on the Ring},
  pdfauthor={L. Forbes, J. Grossman, M. Avery, R. Goh, and G.K. Ocker}
}
\fi

\begin{document}
\maketitle

% REQUIRED
\begin{abstract}
Spatiotemporal neural activity patterns are key features of sensory perception, decision-making, and working memory.
These patterns depend on both the structure of synaptic connectivity and the intrinsic dynamics of neurons and synapses.
Here, we study spatiotemporal dynamics in a simple spiking neuron model: soft-threshold integrate-and-fire networks. 
We extend a recent mean-field theory for these networks to incorporate temporally delayed and spatially nonlocal interactions. The resulting neural field equation resembles the classic Amari-Grossberg model, with an additional term from the reset of the membrane voltage following the emission of an action potential. 
We identify spatial, temporal, and spatiotemporal primary instabilities of homogenous equilibria.
We numerically continue both temporally and spatially periodic solutions of the mean-field equation and track their spectral stability to identify folds of finite-amplitude oscillations, as well as codimension-2 bifurcation points. 
This allows us to provide an initial characterization of the mean-field phase diagram of the system, including regions of bi- and tristability. 
We also identify regimes of complex spatiotemporal dynamics far from the primary instabilities. 
Simulations of the underlying stochastic model confirm all these predictions of the mean-field theory.
\end{abstract}

% REQUIRED

% REQUIRED
\begin{MSCcodes} % https://mathscinet.ams.org/mathscinet/msc/msc2010.html
92B20, 92C20, 35Q70, 35Q92, 35B36, 60G55
\end{MSCcodes}

\section{Introduction}

Spatially and temporally patterned behavior in large-scale neuronal activity are thought to play a role in a wide range of neurobiological phenomena. 
Some examples include geometric visual hallucinations \cite{ermentroutMathematicalTheoryVisual1979}, orientation tuning in the visual cortex \cite{ben-yishaiTheoryOrientationTuning1995}, and short term working memory \cite{camperiModelVisuospatialWorking1998, LaingEtAllMultipleBumpsNeuronal2002}. 
These spatiotemporal dynamics are often understood using neural field equations, which are macroscopic descriptions of large networks' activity \cite{ amariDynamicsPatternFormation1977, bressloffSpatiotemporalDynamicsContinuum2012, grossbergContourEnhancementShort1982, grossbergLearningEnergyentropyDependence1969, wilsonExcitatoryInhibitoryInteractions1972, wilson_mathematical_1973}.
Although they have proven extremely useful in modeling neuronal dynamics, these macroscopic equations lack biophysical detail and can fail to capture spatiotemporal behaviors observed in microscopic spiking network models \cite{byrneNextgenerationNeuralField2019}.
The lack of a direct connection between scales obscures the relation between single-cell and population dynamics.

Integrate-and-fire (IF) models can capture a range of single-cell neural dynamics
\cite{ jolivetQuantitativeSingleneuronModeling2008, naudFiringPatternsAdaptive2008, teeterGeneralizedLeakyIntegrateandfire2018}. 
In these models, the nonlinear dynamics of spiking are replaced with a simple fire-and-reset rule where the membrane voltage returns to a fixed value after the neuron emits a spike.
IF models can be lower dimensional approximations of more bio-physical conductance based models such as the Hodgkin-Huxley model \cite{abbottModelNeuronsHodgkinHuxley1990}.
Networks of IF neurons can exhibit spatial, temporal, and spatiotemporal patterns.
Oscillations can be induced by either synaptic delays or excitatory-inhibitory interactions \cite{brunelFastGlobalOscillations1999, brunelDynamicsSparselyConnected2000, vanvreeswijkWhenInhibitionNot1994, whittingtonInhibitionbasedRhythmsExperimental2000}.
The spatial profile of synaptic projections controls transitions between spatially incoherent and coherent activity \cite{laingStationaryBumpsNetworks2001a, rosenbaumBalancedNetworksSpiking2014}.
Despite the simplicity of the single-neuron dynamics, large networks of IF neurons are high dimensional and complex.
% This makes their utility in studying macroscopic population dynamics difficult.
Therefore, to analytically study temporal and spatial transitions in the network-level dynamics, there has been much interest in field theories that are both analytically tractable and directly connected to the microscopic single-neuron properties, e.g., \cite{byrneNextgenerationNeuralField2019, montbrioMacroscopicDescriptionNetworks2015, schwalgerMindLastSpike2019}.

Here, we study soft-threshold leaky IF (sLIF) networks using a mean-field theory
\cite{ockerDynamicsStochasticIntegratefire2023}.
In particular, we explore oscillatory, spatial, and spatiotemporal instabilities in networks of sLIF neurons with temporal or spatial structure in their synaptic interactions.
First, we introduce the sLIF network and mean-field approximation, and calculate a dispersion relation to identify various instabilities (Section \ref{sec: Homogeneous Equilibria and the Dispersion Relation}).
These instabilities, along with bifurcations of the resulting patterns, delineate regions in the parameter space phase diagram where there exist one or more qualitatively distinct steady state solutions of the mean-field theory.
Each of these steady states correspond to a different type of behavior exhibited by the network.

Taken together, our findings reveal how different structures of connectivity and temporal dynamics give rise to a rich repertoire of network behaviors in a spatially extended sLIF network. 
In particular, we show that transmission delays and spatially structured excitation-inhibition in the synaptic connections support multiple kinds of activity states across different network configurations, as well as multi-stability within individual ones, and we further examine analyze the effect of more complex temporal kernels on the emergence of instabilities (Appendix~\ref{sec: More Realistic Synaptic Temporal Responses}).
We identify and analyze four primary types of behavior and ten distinct regions of parameter space:
\vspace{2mm}

\noindent
(Sections~\ref{sec: Homogeneous Equilibria and the Dispersion Relation}–\ref{sec: Homogeneous Instabilities - Saddle-Node Bifurcation}) \textit{Homogeneous steady states}, constant in both time and space, 

corresponding to globally active or quiescent network states,
\begin{itemize}
\item[(i)] Quiescent region,
\item[(ii)] High activity region,
\item[(iii)] Quiescent and high activity region,
\end{itemize}
\vspace{2mm}
\noindent
(Section~\ref{sec: Delay Induced Hopf Instability}) \textit{Homogeneous oscillatory states}, 
corresponding to global oscillations in network activity,
\begin{itemize}
\item[(iv)] Oscillatory region,
\item[(v)] Oscillatory and high activity region,
\end{itemize}
\vspace{2mm}
\noindent
(Section~\ref{sec: Spatial Instabilities - Turing Bifurcation})
\textit{Stationary spatial patterns}, in which only a portion of the network is above threshold, 
corresponding to spatially patterned activity,
\begin{itemize}
\item[(vi)] Spatially patterned activity region,
\item[(vii)] Spatially patterned and high activity region,
\item[(viii)] Quiescent and spatially patterned activity region,
\item[(ix)] Spatially patterned, quiescent, and high activity region,
\end{itemize}
\vspace{2mm}
\noindent
(Section~\ref{sec: spatiotemporal Instabilities- Turing-Hopf Bifurcation})
\textit{Spatiotemporal patterns}, including standing waves, traveling waves, and oscillating Turing patterns,
\begin{itemize}
\item[(x)] Standing and traveling wave region.
\end{itemize}

\section{The model}
\label{Sec: The model}
\begin{figure}[h]
    \centering
    \includegraphics[width=\textwidth]{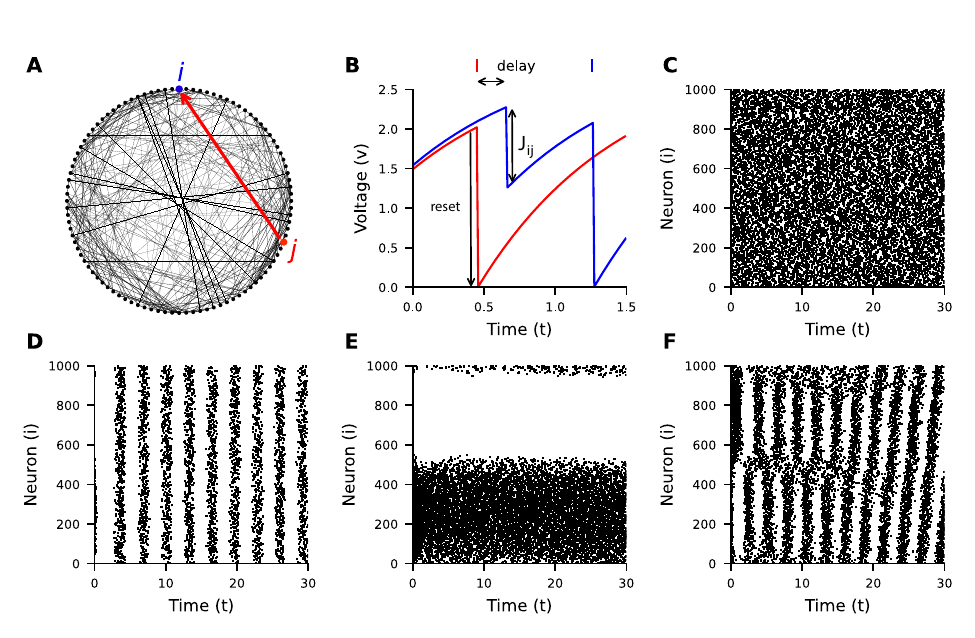}
    
    \caption{Activity patterns in a sLIF network. In all panels, we use the coupling (\eqref{Jspacetime}-\eqref{cosinecoupling}) $J_{ij}(t-D) = \frac{1}{2\pi} \delta(t-D) (J_0 + J_1 \cos(|x_i-x_j|))$ and parameters $D=1$, $E=2$, and $N=1000$. 
    \textbf{A)} A connected network on the ring with a pre and postsynaptic neurons $j$ and $i$.
    \textbf{B)} The delayed impact of a spike from a presynaptic neuron (red) on a postsynaptic neuron (blue). Spike times are marked with tick marks above. $J_0=-1$. 
    \textbf{C)} Homogeneous high activity state, $J_0=-2$, $J_1 = 0$.
    \textbf{D)} Homogeneous oscillatory activity, $J_0=-15$, $J_1=0$.
    \textbf{E)} Spatially patterned activity, $J_0=-2$, $J_1=20$.
    \textbf{F)} A standing wave transitioning into a traveling wave, $J_0=-2$, $J_1=-20$.
    \label{fig:delay}
    }
\end{figure}

We use a soft-threshold leaky integrate-and-fire neuron (sLIF) as our microscopic model to describe how individual neurons integrate incoming signals and generate action potentials (or spikes).
We consider a network of $N$ sLIF neurons with sparse random connectivity on the ring (Fig.~\ref{fig:delay}A).
One-dimensional networks defined on periodic domains have been used to model feature selectivity in neuronal populations---for example, the sensitivity of neurons in the visual cortex to the orientation of visual stimuli \cite{ben-yishaiTheoryOrientationTuning1995, hansel199813}.
Neuron $i$ at location $x_i\in[-\pi,\pi)$ has membrane voltage $v_{i}(t)$ at time $t$. 
Let $n_{j}(t)$ be the counting process associated with neuron $j$, recording the total number of spikes that neuron $j$ has emitted up until time $t$. The increments of that process are $dn_{j}(t)$. After neuron $j$ emits a spike, its membrane potential is reset to the fixed value $r$. 
% At each time step, neuron $j$ is either silent or generates a spike, ($dn_{j,t}\in\{0,1\}$). 
A postsynaptic neuron $i$ receives and integrates spikes from presynaptic neuron $j$ through the synaptic filter $J_{ij}(s-D)$, where $s$ is the time since the presynaptic spike and $D$ the synaptic delay (Fig.~\ref{fig:delay}B).

We implement stochastic spike emission to model the variability in the membrane potentials at which neurons emit a spike \cite{gerstnerPopulationDynamicsSpiking2000}.
This soft threshold for spike emission distinguishes the sLIF model from classic LIF models, where neurons deterministically emit a spike at when the voltage hits a threshold value from below.
The increments $dn_j(t)$ define a point process, $\dot{n}_j(t)$, also called the spike train. Its intensity is given by a non-negative function, $f(v_j(t))$.
%Let $dn_{i,t}$ be modeled by a Bernoulli random variable with probability $f(v_{i,t})dt$, where $f$ is an intensity function dependent on the current membrane potential. 
The intensity function is often chosen to be zero below a threshold, so there is no chance of emitting a spike at low membrane voltages. 
We take $f(v_j)$ to increase unboundedly with $v_j$, so that if the voltage blows up the neuron will be guaranteed to emit a spike and reset.
%Above threshold, $f$ is increasingly higher so there is an increasing chance of generating a spike at higher voltages.

Finally, each neuron has resting potential $E_i$, also incorporating any external input currents. 
Together, the membrane voltage of neuron $i$, $v_{i}(t)$, evolves according to the It\^{o} stochastic differential equation
\begin{equation} \label{discreteLIF}
dv_{i}(t) = \frac{dt}{\tau}\left(-v_{i}(t) + E_i + \frac{1}{N} \sum_{j=1}^N \int J_{ij}(t-s-D) \, dn_{j}(s) \right) - dn_{i}(t^+)\left(v_{i}(t)-r \right)
\end{equation}
% Thus, the membrane voltage of neuron $i$, $v_{i,t}$ evolves according to
% \begin{equation}
%     dv_{i,t} = \frac{dt}{\tau}(-v_{i,t}+E_i+\sum_{k=1}^t\sum_{j=1}^N J_{ij,t-k}dn_{j,k}) - dn_{i,t}(v_{i,t}-r)
%     \label{discreteLIF}
% \end{equation}
% where $k$ indexes all previous time steps, $j$ indexes all neurons in the network, and $\tau$ is the membrane time constant.

%Having the variable $n$ for the spikes allows us to incorporate the reset explicitly into the differential equation \eqref{discreteLIF}.

%Next, we take the continuous time and space limits ($N\rightarrow\infty$ and $dt\rightarrow 0$) of \eqref{discreteLIF}, and assume weak coupling ($J_{ij,t}\sim2\pi/N$). 
%In the continuous-time limit, the Bernoulli process $d{n}_{i,t}$ becomes a point process $\dot{n}(x,t) = \partial_t n(x,t)$ with intensity $f(v(x,t))$ and the synaptic filters $J_{ij,t}$ limit to the function $J(x,t)$. 

The neuron integrates input spikes from the rest of the network through the synaptic filter, or coupling function, $J_{ij}(t)$.
We show that with simple pulse coupling and no synaptic delay, $J_{ij}(t)=J_0\delta(t)$, the sLIF network can exhibit homogeneous spiking behavior (Fig.~\ref{fig:delay}C).
When $J$ has more temporal and spatial structure, the network exhibits spatiotemporal dynamics.
%We assume the coupling kernel satisfies $J(x,t)\in L^1([-\pi,\pi)\times\mathbb{R}_+)$ with separate temporal and spatial profiles of the synaptic current.
%For the temporal structure, we assume a causal kernel with an explicit time delay $D$.
%This models transmission delays of spikes between pre- and postsynaptic neurons (Fig.~\ref{fig:delay}B, red and blue).
For non-zero delay values, the sLIF network can exhibit homogeneous oscillatory activity (Fig.~\ref{fig:delay}D).
Additionally, with sufficient spatial modulation in $J$, the sLIF network exhibits spatially patterned activity (Fig.~\ref{fig:delay}E). 
Finally, if our coupling function has both delays and spatial modulation, the network can exhibit spatiotemporal dynamics with patterns of traveling or standing waves (Fig.~\ref{fig:delay}F, which also shows a spontaneous, noise-induced transition between those two patterns). 
To understand these dynamics, we develop a neural field approximation of \eqref{discreteLIF}.

\subsection{Mean-field theory}
We will study the following mean-field theory for \eqref{discreteLIF},
\begin{equation}
    \partial_t \bar{v}(x, t) = -\bar{v}(x, t) + E + \int_{-\infty}^{t-D} \int_{-\pi}^\pi J(x-y, t-s-D) \, \bar{f}(y, s) \, dy \, ds \, - \bar{f}(x, t) \, \bar{v}(x, t)
    \label{mean-field}
\end{equation}
Here, $\bar{v}(x,t)$ is the mean membrane potential on the ring $x\in[-\pi,\pi)$, $\bar{f}:=f(\bar{v})$ is the mean-field approximation of the spike trains $\dot{n}$, $J \in L^1([\pi, \pi) \times \mathbb{R}_+$) is the synaptic filter, and $E$ is the mean resting potential. 
We have non-dimensionalized the model by measuring time relative to $\tau$ and the membrane potential relative to $r$, setting $\tau=1$ and $r=0$. 
The presence of the reset term $- \bar{f}(x, t) \bar{v}(x,t)$ differentiates this model from the classic Amari-Grossberg activity equations \cite{amariDynamicsPatternFormation1977, grossbergLearningEnergyentropyDependence1969}. 

Rigorous mean-field limits for the population density of soft-threshold integrate-and-fire networks with delays, spatial structure, and disordered connectivity have each been studied separately, e.g.,~\cite{chevallier_mean-field_2017, cormierHopfBifurcationMeanField2021, duarte_hydrodynamic_2015, jabinDenseNetworksIntegrateandfire2024, jabin_mean-field_2025, robertDynamicsRandomNeuronal2016}. To our knowledge, a mean-field limit for soft-threshold integrate-and-fire networks with all those factors combined has not yet been rigorously established. 
Here, we instead study a simple approximate mean-field theory for the expected voltage \eqref{mean-field}, which can be derived following the method of~\cite{ockerDynamicsStochasticIntegratefire2023}. Even in the continuum limit, $N \to \infty$, \eqref{mean-field} will not be an exact description of the microscopic model; it results from truncating a cumulant hierarchy. We leverage its simplicity, however, for a qualitative understanding of the instabilities of stationary homogeneous states and the resulting activity patterns.

\section{Results}
We begin our study of the dynamics of \eqref{mean-field} by characterizing its homogenous equilibria and their stability. Throughout, we will take a rectified linear intensity function, $f(v)=\lfloor v-1 \rfloor_+$. 

\subsection{Homogeneous equilibria and dispersion relation}
\label{sec: Homogeneous Equilibria and the Dispersion Relation}

There are three possible homogeneous equilibria of \eqref{mean-field}: a quiescent state below threshold ($v_Q$) and two solutions above threshold ($v_\pm$). These are given by
\begin{equation}
v_0 =
\begin{cases}
v_Q = E & \text{if } E < 1 ,
\\[2 pt]
v_+ =  \left( J_0 + \sqrt{J_0^2 + 4(E-J_0)}\right)/2 & \text{if } J_0>2+2\sqrt{1-E} \text{ or }J_0<2,E>1,
\\[6 pt]
v_- =  \left( J_0 - \sqrt{J_0^2 + 4(E-J_0)} \right)/2 & \text{if } J_0>2+2 \sqrt{1-E} \text{ and }  J_0>2, E<1 ,
\end{cases}
\label{homog_solns}
\end{equation}
where $J_0=\int \int J(x-y,t-s)\, ds \, dy$.

The quiescent sub-threshold state $v_Q$ is always stable. Below threshold, the convolution and reset terms of \eqref{mean-field} vanish, and the voltage converges to the resting potential $E$.
The supra-threshold state $v_+(v_-)$ is stable(unstable) in absence of delay and spatial modulation in the coupling.
Therefore, without spatial or temporal structure in the coupling, the mean activity will converge to $v_Q$ or $v_+$ \cite{ockerDynamicsStochasticIntegratefire2023}. 
The first three regions we identify in the phase diagram (Fig.~\ref{fig:deltaStochVsMean}A), where the stable homogeneous steady states exist, are: \\

\begin{enumerate}
\item[i)] Q: Only a quiescent state ($v_Q$, below threshold) exists and is stable.
\item[ii)] H: Only a high activity state ($v_+$, above threshold) exists and is stable.
\item[iii)] Q-H: Both the quiescent and high activity states are stable. \\
\end{enumerate} 

In the bistable region Q-H, delimited by saddle-node bifurcations (Section \ref{sec: Homogeneous Instabilities - Saddle-Node Bifurcation}), the unstable equilibrium $v_-$ is the separatrix between $v_Q$ and $v_+$.

To identify the instabilities of a homogeneous state $v_0$, we perform a linear stability analysis of the mean-field approximation \eqref{mean-field}, which yields a dispersion relation characterizing the stability of $v_0$ with respect to perturbations in each spatial Fourier mode. 
Given the threshold-linear intensity function, $v_0$ above threshold ($v_0>1$), and recalling the delay $D$ in the synaptic coupling, the linearization of \eqref{mean-field} about $v_0$ is
\begin{equation}
    \partial_t w(x, t) = -2 v_0 w(x, t) + \int_{-\infty}^{t-D} \int_{-\pi}^\pi J(x-y, t-s-D) \, w(y,s)\, dy\,ds
    \label{linearization}
\end{equation}
By taking the spatial Fourier transform of \eqref{linearization} $\left(w(x,t)=\sum_{k\in\mathbb{Z}}\hat{w}_k(t)e^{ikx}\right)$ and making the ansatz $\hat{w}_k(t)=e^{\lambda_k t}$, we obtain the dispersion relation
\begin{equation}
    \lambda_k = -2v_0+ e^{-\lambda_k D}\int_{0}^\infty \hat{J}_k(s) \, e^{-\lambda_k s}\,ds,\text{ }k\in\mathbb{Z}
    \label{general dispersion1}
\end{equation}
where $\lambda_k$ is the eigenvalue associated with the $k^{th}$ mode, $\hat{w}_k$.
The stability depends on the structure of interactions between the neurons through the time-dependent Fourier coefficients of the coupling function $\hat{J}_k(t)=\int e^{-ikx}J(x,t)\,dx$.
The dispersion relation for a general intensity function can be found in Appendix \ref{sec: General Dispersion Relationship}.

For simplicity, we assume delayed pulse and cosine coupling for the temporal and spatial structure of the coupling function:
\begin{align}
J(x, t-D) =&  J_{\mathrm{space}}(x) \, J_{\mathrm{time}}(t-D), \label{Jspacetime}\\
J_{\mathrm{time}}(t-D)=&\delta(t-D), \label{delayedpulse} \\
J_{\mathrm{space}}(x)=&\frac{1}{2\pi}\left(J_0 + J_1\cos(x)\right). \label{cosinecoupling}
\end{align}
% \begin{equation}
%     J_{\mathrm{time}}(t-D)=\delta(t-D)
%     \label{delayedpulse}
% \end{equation}
% for the temporal profile and for the spatial profile we assume cosine coupling
% \begin{equation}
%     J_{\mathrm{space}}(x)=\frac{1}{2\pi}\left(J_0 + J_1\cos(x)\right)
%     \label{cosinecoupling}
% \end{equation}

With delayed pulse coupling, the dispersion relation \eqref{general dispersion1} simplifies to
\begin{equation}
    \lambda_k = -2v_0 + \hat{J}_ke^{-\lambda_k D},
    \label{delta_dispersion}
\end{equation}
where $\hat{J}_k$ is the $k^{th}$ Fourier coefficient of the spatial coupling. Given cosine coupling \eqref{cosinecoupling}, there are only three nonzero Fourier coefficients, $\hat{J}_0 = J_0$ and $\hat{J}_{\pm1} = J_1/2$.
Therefore, all spatial modes other than $k=0$ and $k=1$ are always stable.

The analysis above extends to other coupling structures.
Other $L^2$ spatial kernel functions can be studied in a similar way by substituting their Fourier coefficients into \eqref{general dispersion1}. 
For other choices of temporal kernels, the temporal integral in \eqref{general dispersion1} will differ.
We also analyze two other temporal kernels which model the rise and fall times of the action potential in addition to the delay (Appendix~\ref{sec: More Realistic Synaptic Temporal Responses}).

The dispersion relation \eqref{delta_dispersion} only pertains to equilibria above threshold. 
As discussed above, the only sub-threshold homogeneous equilibria $v_Q$ is always stable. 
Consequently, we restrict our analysis of instabilities to the two steady states above threshold ($v_\pm$). 
The location of primary instabilities are given by the dispersion relation \eqref{delta_dispersion} when Re$(\lambda_k)=0$ for different modes $k$ and temporal frequencies Im$(\lambda_k)$.

These instabilities correspond to four smooth bifurcations of $v_0$: a saddle-node bifurcation (Section \ref{sec: Homogeneous Instabilities - Saddle-Node Bifurcation}), a Hopf bifurcation (Section \ref{sec: Delay Induced Hopf Instability}), a Turing bifurcation (Section \ref{sec: Spatial Instabilities - Turing Bifurcation}), and a Turing-Hopf bifurcation (Section \ref{sec: spatiotemporal Instabilities- Turing-Hopf Bifurcation}). 
At threshold ($v_0=1$), non-smooth bifurcations can occur due to the non-smooth point of $f$, not described by the linearized analysis; we therefore study these numerically.

\subsection{Homogeneous, static instabilities}
\label{sec: Homogeneous Instabilities - Saddle-Node Bifurcation}
\begin{figure}[htbp]
    \centering
    \includegraphics[width=\textwidth]{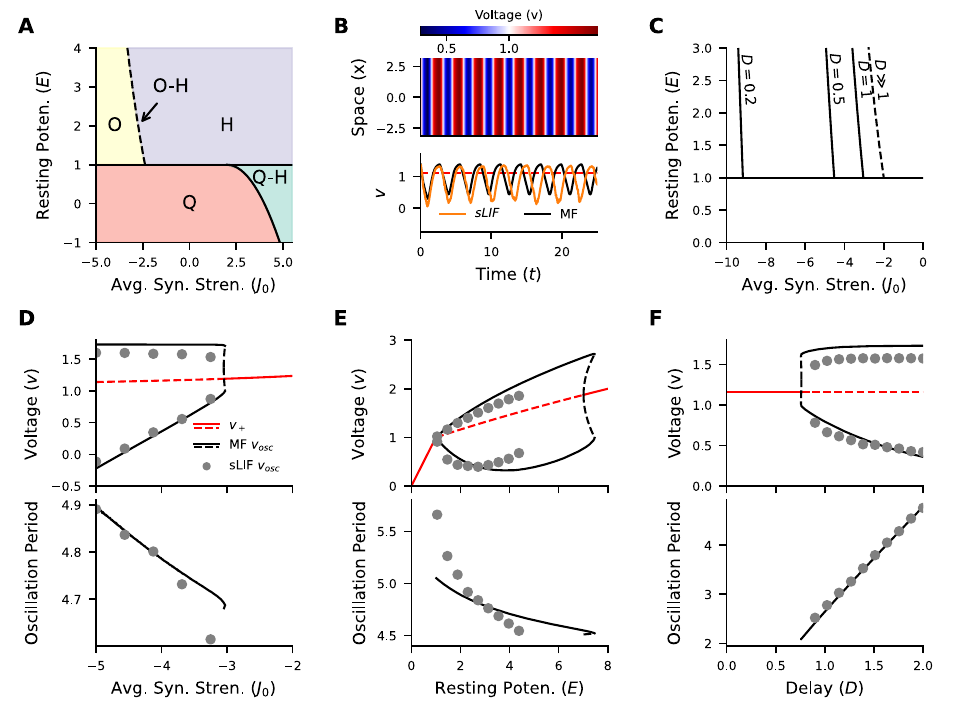}
    \caption{\label{fig:deltaStochVsMean} Homogeneous oscillations in the mean-field approximation \eqref{mean-field} and the sLIF Network \eqref{discreteLIF} with $N=1000$ and delayed pulse coupling \eqref{delayedpulse}. Unless otherwise stated, parameters are $J_0=-4$, $J_1=0$, $E=3$, and $D=2$. 
    \textbf{A)} Phase diagram of the mean-field approximation \eqref{mean-field} in the $J_0,E$ plane. Quiescent activity (Q, red), high activity (H, purple), bistable region of quiescent and high activity (Q-H, green), oscillatory (O, yellow), and bistable region of oscillatory and high activity (O-H, dashed black line).  
    \textbf{B)} Upper: homogeneous oscillation in mean-field approximation. Lower: The comparison of the population averages of both the above mean-field (black) and the sLIF network (orange) simulated at the same parameters
    \textbf{C)} Hopf curves \eqref{deltaHopfEq} in the $J_0,E$ plane with $D=\{0.2, 0.5, 1\}$. The dashed line shows the location of the Hopf curve as $D\rightarrow\infty$.
    \textbf{D–F)} Bifurcation diagrams (\textit{upper}) and steady-state oscillation periods (\textit{lower}) varying $J_0$, $E$, and $D$, respectively. In the upper panels, solid (dashed) black lines denote the min/max amplitude of stable (unstable) limit cycles; red lines indicate the homogeneous solution. Grey circles show sLIF simulation data.
    }      
\end{figure}

The first instability of the homogeneous state occurs at $k=0$ with Im($\lambda_{0}) =0$. 
Setting Re$(\lambda_0)=0$ in the dispersion relation \eqref{delta_dispersion} gives the saddle-node bifurcation where $v_\pm$ collide and vanish,
\begin{equation}
J_0=2+2\sqrt{1-E},
\end{equation}
where $J_0>2$ and $E<1$.
Beyond the bifurcation, at lower $J_0$ values, solutions tend toward the quiescent state $v_Q$. 
The saddle-node curve forms the boundary between the bistable regime (Q-H) and the quiescent regime (Q) (Fig.~\ref{fig:deltaStochVsMean}A).
This bifurcation is independent of the delay and spatial coupling.

There is also a non-smooth saddle-node bifurcation at $E=1$ where $v_-$ and $v_Q$ collide and vanish.
This bifurcation cannot be identified via the dispersion relation, which is derived from the linearized dynamics above threshold. 
The non-smooth saddle-node curve is the boundary between the bistable region (Q-H) and high activity region (H) (Fig.~\ref{fig:deltaStochVsMean}A).
These bifurcations were identified in \cite{ockerDynamicsStochasticIntegratefire2023}.
If the intensity function $f(v)$ is not threshold-linear, another saddle-node bifurcation between high and low activity states can emerge \cite{paliwalMetastabilityNetworksStochastic2025}.

\subsection{Delay-induced instability}
\label{sec: Delay Induced Hopf Instability}

Transmission delays can induce a Hopf bifurcation, at which homogeneous (bulk) oscillations of the network emerge. 
This oscillatory instability occurs at $k=0$ with Im($\lambda_{0}) \neq0$ when a pair of complex conjugate eigenvalues cross the imaginary axis ($\mathrm{Re}(\lambda_0)=0$).
From the dispersion relation \eqref{delta_dispersion}, we find that $v_+$ undergoes a Hopf bifurcation when
\begin{equation}
    \frac{-1}{D}\cos^{-1}\left(\frac{2v_+}{J_0}\right)  = J_0 \sqrt{1 - \left(\frac{2v_+}{J_0}\right)^2}
    \label{deltaHopfEq}
\end{equation}
and $v_+>1 $, $3J_0^2+4J_0 > 4E$, $D>0$, and $J_0<0$ (Appendix \ref{sec:Instabilities with Delayed Pulse Coupling}).
Beyond the Hopf bifurcation, at lower $J_0$ values, there are homogeneous oscillatory solutions $v_{osc}$ (Fig.~\ref{fig:deltaStochVsMean}B, upper) if the delay is sufficiently large (Fig.~\ref{fig:deltaStochVsMean}C).
Thus uniform oscillations require inhibitory coupling with sufficient delay and the reversal potential $E$ above threshold. 
However, an excessively large $E$ offsets the inhibition and oscillations do not occur.

We find, through numerical continuation of $v_{osc}$ (Appendix \ref{sec: Numerical Methods}), that the Hopf bifurcation \eqref{deltaHopfEq} is subcritical (Fig.~\ref{fig:deltaStochVsMean}D-F).
Shortly after the minimum voltage on branch of unstable oscillations emerging from the bifurcation falls below threshold, there is a fold of large amplitude limit cycles, where the unstable branch meets a stable branch.
The Hopf bifurcation and the fold of limit cycles bound a narrow bistable region of both $v_+$ and $v_{osc}$ (Fig.~\ref{fig:deltaStochVsMean}A, ``O-H": dotted line). 
As the magnitude of $J_0$ or $E$ decrease or $D$ increase from the their values at the fold, so does the amplitude and the period of the oscillation (Fig.~\ref{fig:deltaStochVsMean}D-F).
As, $E$ further decreases, $v_{osc}$ vanishes at a supercritical non-smooth Hopf bifurcation at $E=1$, where $v_+$ drops below threshold (Fig.~\ref{fig:deltaStochVsMean}E).

%In O-H, the unstable limit cycle between the homogeneous and oscillatory solutions is the boundary between their two basins of attraction.
To summarize, we have found with the presence of delays, there are two additional regions of the phase diagram (Fig.~\ref{fig:deltaStochVsMean}A) involving $v_{osc}$: \\

\begin{enumerate}
\item[iv)] O: Only a homogeneous, large-amplitude oscillation ($v_{osc}$) around $v_+$ is stable.
\item[v)] O-H: Both a homogeneous oscillation ($v_{osc}$) and high activity ($v_+$) state are stable. \\
\end{enumerate}

% \vspace{0.25cm}
% iv) Oscillatory Region (O) - Only a homogeneous large amplitude oscillation around $v_+$ is stable

% v) Oscillatory \& High Activity Region (O-H) - Both a homogeneous oscillation and $v_+$ itself are stable
% \vspace{0.25cm}
\noindent
The Hopf curve \eqref{deltaHopfEq} is the boundary between the oscillatory (O) and bistable (O-H) regions, while the nearby fold of limit cycles is the boundary between the bistable (O-H) and high (H) activity regions. 
The non-smooth supercritical Hopf bifurcation at $E=1$ separates the oscillatory (O) and quiescent (Q) regions.

\begin{figure}[htbp]
    \centering
    \includegraphics[width=\textwidth]{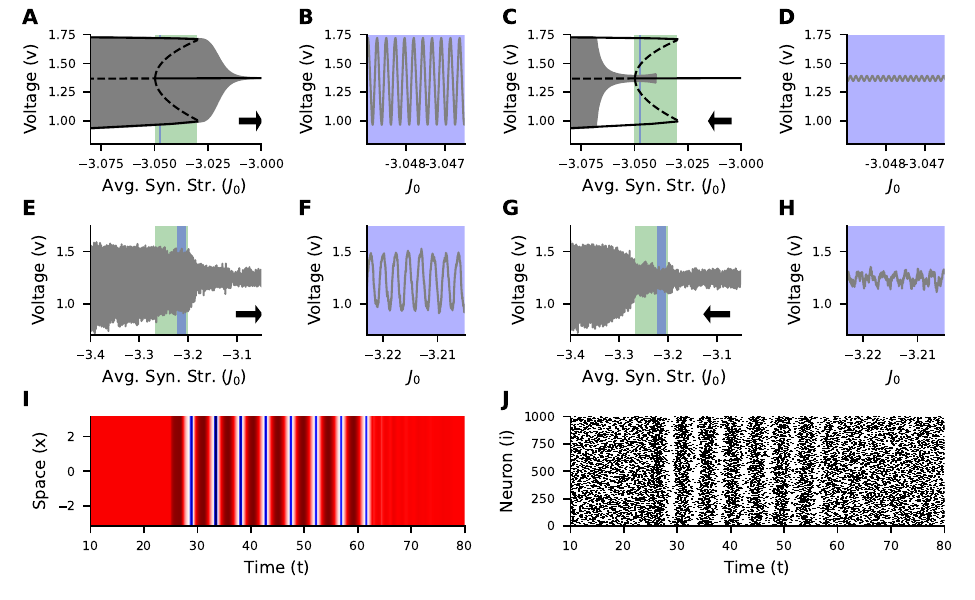}  
    \caption{\label{fig:deltaStochVsMean_bistab}
    Bistability between high activity ($v_+$) and oscillatory states ($v_{osc}$) with $E=3$ and $D=2$. 
    \textbf{A)} The loss of stability of $v_{osc}$ in the mean-field \eqref{mean-field} (grey) at the fold of limit cycles and upper boundary of the bistable region (green) under slow ramp up of $J_0$ (right arrow). 
    The voltage is plotted against the current value of \( J_0 \), overlaid with the bifurcation diagram (black).
    \textbf{B)} A zoom in on the stable oscillation in the bistable region of the mean-field simulation (A, blue) 
    \textbf{C)} Loss of stability of $v_+$ \eqref{mean-field} (grey) at the Hopf bifurcation and lower boundary of the bistable region (green) under slow ramp up of $J_0$ (left arrow).
    \textbf{D)} A zoom in on the near homogeneous solution in the bistable region of the mean-field simulation (C, blue) in the same parameter range as (B).   
    \textbf{E)} Same as (A) but with the sLIF network \eqref{discreteLIF}.
    \textbf{F)} A zoom in on the stable oscillation in the bistable region of the sLIF network simulation (E, blue)
    \textbf{G)} Same as (C) but with sLIF network.
    \textbf{H)} A zoom in on the near homogeneous solution in the bistable region of the sLIF network simulation (G, blue) in the same parameter range as (F).
    \textbf{I)} Homogeneous and oscillatory behavior of the mean-field equation at set parameters ($J_0=-3.04$). Pulses of amplitude 1 and -1 and duration 2 and 1.75 were applied to $E$ at $t=25$ and $t=62.5$ to turn on(off) the oscillatory pattern. 
    \textbf{J)} Homogeneous and oscillatory behavior in the spiking network at set parameters ($J_0=-3.22$). Pulses of amplitude 1 and -1 and duration 2 and 1 were applied to $E$ at $t=25$ and $t=58$.
    }      
\end{figure}

Oscillatory behavior in the sLIF network (Fig.~\ref{fig:deltaStochVsMean}B lower panel, orange) qualitatively matches the mean-field approximation (Fig.~\ref{fig:deltaStochVsMean}B lower panel, black), with some discrepancy expected due to the mean-field approximation.
The maximum amplitude (Fig.~\ref{fig:deltaStochVsMean}D-F) of the sLIF network oscillations (grey circles) are generally lower than those of the mean-field approximation (solid black line). 
Moreover, the oscillatory region of the sLIF network is narrower: the onset of oscillation at the Hopf bifurcation occurs at a lower value of $E$ than predicted by the mean-field theory, while the non-smooth Hopf bifurcation remains at threshold (Fig.~\ref{fig:deltaStochVsMean}E).
The sLIF simulation data are shown only for large amplitude oscillations, which we further investigate in Fig.~\ref{fig:deltaStochVsMean_bistab}.

We observe the small bistable region O-H in the sLIF network with a slow ramp of the bifurcation parameter through the region, starting on either side.
We show this process first with the mean-field approximation, where we know the location of the Hopf bifurcation analytically and the location of the fold of limit cycles from numerical continuation.
We then repeat the process using the sLIF network and observe similar phenomena. 

We initialize a simulation of the mean-field (Fig.~\ref{fig:deltaStochVsMean_bistab}A, grey) in the oscillatory region (O) at a stable limit cycle and slowly increase the parameter $J_0$, through the bistable region (Fig.~\ref{fig:deltaStochVsMean_bistab}A, green).
As the ramp continues, the limit cycle (Fig.~\ref{fig:deltaStochVsMean_bistab}B, Fig.~\ref{fig:deltaStochVsMean_bistab}A blue) persists until it loses stability at the fold marking the end of bistable region.
The network activity then transitions to the stable homogeneous state and the oscillation decays. 

For the reverse ramp (Fig.~\ref{fig:deltaStochVsMean_bistab}C), we choose initial conditions above the bistable region near the stable homogeneous state. 
As $J_0$ is slowly decreased through the bistable region, the network stays close to the stable homogeneous solution (Fig.~\ref{fig:deltaStochVsMean_bistab}D), until it loses stability at the Hopf bifurcation. 
A temporal oscillation then begins to grow and converges to the stable limit cycle.
The delayed loss of stability, where the system stays close to the unstable homogeneous state after the bifurcation point, is expected when simulating smooth dynamical systems near a Hopf bifurcation \cite{izhikevichDynamicalSystemsNeuroscience2014}.
It is due to the time it takes for the system to diverge from the unstable equilibrium.

We repeat both parameter ramps with the sLIF network.
When increasing $J_0$ from the oscillatory region (Fig.~\ref{fig:deltaStochVsMean_bistab}E), we observe a stable oscillation persisting in the bistable region (Fig.~\ref{fig:deltaStochVsMean_bistab}F) and a similar loss of stability of the limit cycle, but at a slightly lower $J_0$ value than that of the mean-field approximation.
Decreasing $J_0$ from the homogeneous regime, we again observe similar behavior to the mean-field (Fig.~\ref{fig:deltaStochVsMean_bistab}G) where the behavior stays close to the homogeneous solution in the bistable region (Fig.~\ref{fig:deltaStochVsMean_bistab}H), and then loses stability, but slightly lower $J_0$ values than the mean-field.

In the sLIF network, the loss of stability of the homogeneous solution occurs at a lower parameter value than where the limit cycle loses stability. 
There is thus a parameter range where the stationary and oscillatory solutions coexist.
We also demonstrate this bistability at set parameter values where a stimulus switches the network between the homogeneous and oscillatory states in both the mean-field approximation (Fig.~\ref{fig:deltaStochVsMean_bistab}I) and the sLIF network (Fig.~\ref{fig:deltaStochVsMean_bistab}J).
%We conclude that there is a subcritical Hopf bifurcation in the spiking network as predicted by the mean-field approximation.

\subsection{Spatial instabilities}
\label{sec: Spatial Instabilities - Turing Bifurcation}

Next, we investigate the instabilities induced by spatial modulation of the coupling function.
A spatial instability occurs at $k\neq0$ with Im$(\lambda_k)=0$ when Re$(\lambda_k)=0$.
Recall that the amount of the spatial modulation of the synaptic coupling is given by the parameter $J_1$.
Setting $k=1$ and $\lambda_1=0$ in the dispersion relation \eqref{delta_dispersion} with cosine coupling \eqref{cosinecoupling} reveals a spatial instability at
\begin{equation}
    4v_0=J_1, \qquad J_1>4.
    \label{Turing}
\end{equation} 

\begin{figure}[h!]
    \centering
    \includegraphics[width=\textwidth]{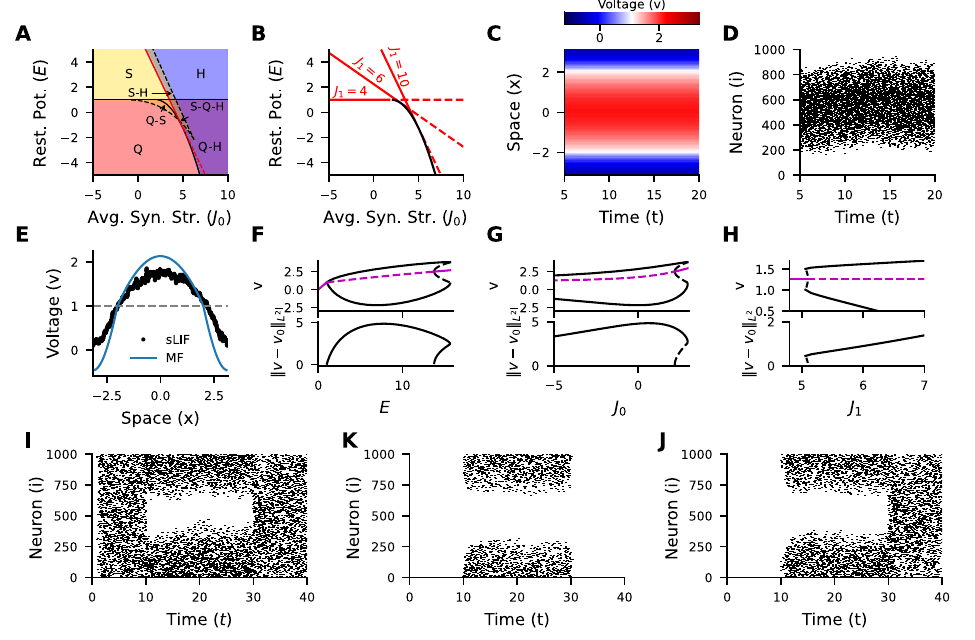} 
    \caption{\label{fig:spatial}
    Spatial patterns in the mean-field approximation \eqref{mean-field} and the sLIF Network \eqref{discreteLIF} with $N=1000$ and cosine coupling \eqref{cosinecoupling} and no delay.
    \textbf{A)} Phase diagram in $J_0,E$ plane showing quiescent (Q, red), high (H, blue), and spatially patterned (S, yellow) activity regions. Bistable regions notated with a hyphen (e.g. Q-H).
    \textbf{B)} Turing bifurcation of $v_+$ (solid red), the spatial instability of $v_-$ (dashed red), and saddle-node bifurcation (black), shown for \( J_1 = \{4,6,10\} \). 
    \textbf{C)} Stable Turing pattern solution of the mean-field model at $E = 3$, $J_0 = -2$, $J_1 = 8$.  
    \textbf{D)} Corresponding Turing pattern in a spiking network with $N = 1000$.  
    \textbf{E)} the comparison of the solution profiles of mean-field (blue) and sLIF network (black) from the simulations shown in (C) and (D) at $t=20$, smoothed over time and space, shown with the firing threshold (grey dashed).   
    \textbf{F-H)} Bifurcation diagrams (top) showing the subcritical Turing bifurcation: Turing pattern amplitude (black lines, min/max) and homogeneous states (purple), with solid/dashed lines indicating stability. 
    Bottom panels show the $L^2$ norm $||v-v_+||_{L^2(-\pi,\pi)}$ of the Turing patterns. 
    Continuation shown with:
    \textbf{F)} varying $E$ ($J_0 = 5$, $J_1 = 10$), 
    \textbf{G)} varying $J_0$ ($E=5$, $J_1=10$), 
    \textbf{H)} varying $J_1$ ($E=5$, $J_0=5$),
    \textbf{I-K)} Multi-stability in the spiking network:
    \textbf{I} the high activity and spatially patterned states at $J_0=3.1$, $J_1=10$, and $E=1.75$,
    \textbf{J)} the quiescent and spatially patterned states at $J_0=3$, $J_1=0$, and $E=0.75$,
    \textbf{K)} and the quiescent, spatially patterned, and high activity states at $J_0=4.2$, $J_1=10$, $E=0.5$. 
    Transitions between states were induced by pulse inputs to $E$ at $t=10$ and $t=30$.
    }      
\end{figure}

The spatial instability \eqref{Turing} takes one of two forms, depending on $J_0$. For $J_0 < J_1/2$, it is a Turing bifurcation of $v_+$ since the $k=0$ mode remains stable, while for $J_0 > J_1/2$, it is a secondary instability of $v_-$, whose $k=0$ mode is already unstable (Fig.~\ref{fig:spatial}A, solid and dashed red curves).

The spatial instability switches from $v_+$ to $v_-$ at $J_0=J_1/2$ and $E=J_0-J_0^2/4$, where the Turing and saddle-node curves meet tangentially at a codimension-2 bifurcation (Fig. ~\ref{fig:spatial}A, red and black curves).
At this point, the $k=0$ and $k=1$ Fourier coefficients of the coupling function are equal and the saddle-node (instability of the $k=0$ mode) and Turing (instability of the $k=1$ mode) bifurcations coincide.

The existence and location of the spatial instability curve depends on the amount of spatial modulation in the coupling function (Fig.~\ref{fig:spatial}B).
If $J_1 \leq 4$, no spatial instability occurs.
This critical value emerges because at $J_1=4$, the spatial instability curve lies at the boundary of the regions of where $v_\pm$ are above threshold ($E=1$). 
If $J_1<4$, then \eqref{Turing} is only defined in regions where $v_\pm$ are below threshold and no longer correspond to equilibria. 
Therefore $J_1$ must exceed this critical value ($J_1>4$) for pattern formation to occur.
Additionally, the Turing bifurcation of $v_+$ only occurs when the spatial coupling is either purely inhibitory ($J_0<-|J_1|$) or has local excitation and long range inhibition ($|J_0|<J_1$), as illustrated later in Figs. \ref{fig:spatiotemporal}A and B (Section \ref{sec: spatiotemporal Instabilities- Turing-Hopf Bifurcation}).

% \vspace{0.5cm}
% \noindent
% \textit{The Turing pattern Solution--}

\subsubsection{A Turing pattern}
Beyond the spatial instability \eqref{Turing}, the system supports a stationary periodic Turing pattern (Fig.~\ref{fig:spatial}C).
The profile of the Turing pattern has portions above and below threshold, corresponding to the active and inactive portions of the network.
We refer to this as the spatially patterned state (S).
The sLIF network exhibits qualitatively similar spatial patterns (Fig.~\ref{fig:spatial}D,E).

The Turing bifurcation is subcritical with a nearby fold of Turing patterns.
We used numerical continuation of the Turing pattern and determined its stability by numerically computing eigenvalues of the mean-field linearization about the Turing pattern (Appendix \ref{sec: Numerical Methods}).
An unstable branch of small-amplitude Turing patterns (Fig.~\ref{fig:spatial}F-H, dashed black) emerges backward when the active state $v_+$ (Fig.~\ref{fig:spatial}F-H, purple) destabilizes through the Turing bifurcation as $E$ or $J_0$ decrease or $J_1$ increases.
The numerical continuation revealed that all completely supra-threshold Turing patterns are unstable.
The minimum value of the unstable Turing pattern drops below threshold at $v=1$, it soon after meets a stable branch of Turing patterns at a fold.
The stable branch extends off in both $J_0$ (Fig.  \ref{fig:spatial}G) and $J_1$ (Fig.  \ref{fig:spatial}H), but ends at a non-smooth Turing at $E=1$ when $v_+$ drops below threshold (Fig.  \ref{fig:spatial}F).

\subsubsection{Multi-stability with the Turing pattern}
Due to the subcritical nature of the Turing bifurcation, there exists a bistable region of both the homogeneous state and Turing pattern between the bifurcation point and the fold.
Within this region, networks can support either spatially patterned or high activity (Fig.~\ref{fig:spatial}A, ``S-H'').
Additionally, we find that in a small neighborhood of the codimension-2 bifurcation of the Turing and saddle-node, the non-smooth Turing is also subcritical with a nearby fold extending into the quiescent region.
This gives rise to two additional regions of multi-stability with the Turing pattern involving the quiescent state: a quiescent and spatially patterned region (``Q-S'') and a region of triple stability between the quiescent , spatially patterned, and high activity states (``S-Q-H'')  when the subcritical regions of the two bifurcations overlap.
A detailed characterization of the regions and their boundaries, through numerical continuation and tracking the folds, is provided in Appendix \ref{sec: Tracking the fold points to identify the boundaries of bistable regions}.

We have identified four additional regions of the phase diagram (Fig.~\ref{fig:spatial}A) involving the Turing pattern: \\
\begin{enumerate}
\item[vi)] S: A stable spatially patterned solution exists.
\item[vii)] S-H: Spatially patterned and High ($v_+$) activity solutions exist and are stable.
\item[viii)] Q-S: Stable Quiescent ($v_Q$) and spatially patterned solutions exist.
\item[ix)] S-Q-H: Spatially patterned, Quiescent ($v_Q$), and High ($v_+$) activity solutions exist and are stable. \\
\end{enumerate}

The spiking network also exhibits multi-stability in these regions.
We observe bistability between the spatially patterned and high activity states, as predicted by the subcritical Turing bifurcation in the mean-field approximation (Fig.~\ref{fig:spatial}I).
Additionally, we observe the bistability between the quiescent and spatially patterned activity states (Fig.~\ref{fig:spatial}J), as predicted by the subcritical non-smooth Turing, and multi-stability between all three states (Fig.~\ref{fig:spatial}K).

% \vspace{0.25cm}

% vi) Local Activity (L) - A stable Turing pattern exists

% vii) Quiescent or Local Activity (Q-L) - Both $v_Q$ and Turing pattern exist and are stable

% \vspace{0.25cm}

% \vspace{0.5cm}
% \noindent
% \textit{Multi-stability of the Turing pattern and the quiescent state--}
%\subsubsection{Multi-stability of the Turing pattern and the quiescent state}

% \vspace{0.25cm}

% viii) Local or High Activity (L-H) - Both $v_+$ and Turing pattern exist and are stable

% ix) Local, Quiescent, or High Activity (L-Q-H) - All of $v_Q$, $v_+$, and Turing patterns exist and are stable 

% \vspace{0.25cm}

% \vspace{0.5cm}
% \noindent
% \textit{Tracking the fold points--}

\subsection{Spatiotemporal instabilities}
\label{sec: spatiotemporal Instabilities- Turing-Hopf Bifurcation}
In the presence of both a delay and spatial modulation in the coupling, a spatiotemporal instability occurs and gives rise to spatiotemporal patterns.
From the dispersion relation \eqref{delta_dispersion} we find the spatiotemporal instability in the first spatial mode ($k=1$) at nonzero frequency (Im($\lambda_{1})\neq0$). It occurs when $v_0$ is above threshold and
\begin{equation}
    \frac{-1}{D}\cos^{-1}\left(\frac{4v_0}{J_1}\right)  = \frac{J_1}{2} \sqrt{1 - \left(\frac{4v_0}{J_1}\right)^2}.
    \label{deltaTuringHopfEq}
\end{equation}
This instability is a Turing-Hopf bifurcation of $v_+$ if $J_0<J_*$ and a secondary instability of $v_-$ if $J_0>J_*$, where $J_*$ is the $J_0$ value of the codimension 2 point at which the Turing-Hopf curve and saddle-node curve intersect.

\begin{figure}[h!]
    \centering
    \includegraphics[width=\textwidth]{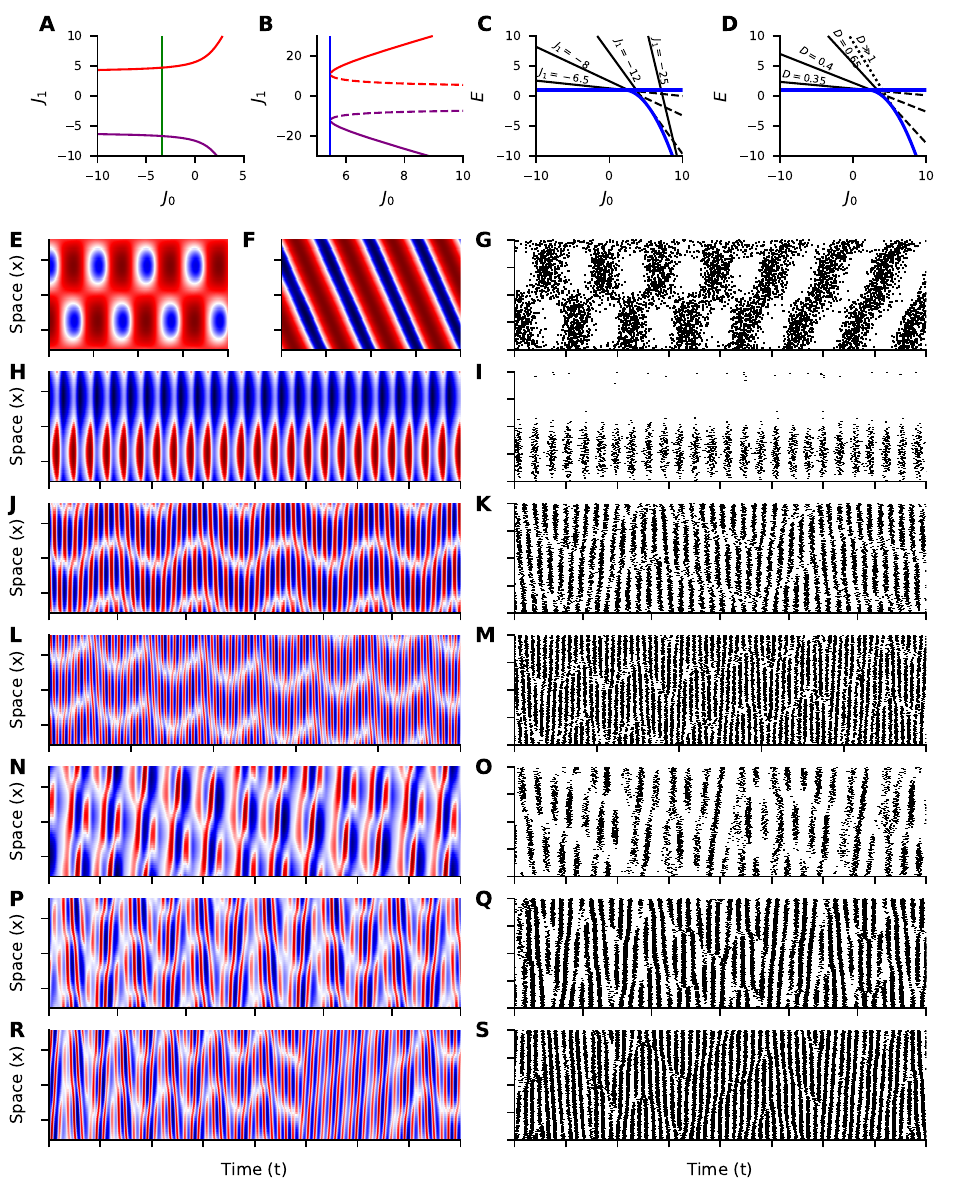}
    \caption{Spatiotemporal instabilities and patterns in the mean-field approximation \eqref{mean-field} and the sLIF network \eqref{discreteLIF}.
    \textbf{A)} Primary instabilities at $E=2$, $D=1$: Turing (red), Hopf (green), and Turing-Hopf (purple).
    \textbf{B)} Same as A, but at $E=-2$: includes Saddle (blue), and Turing-Hopf of $v_+$ (solid) and $v_-$ (dashed).
    \textbf{C-D)} Turing-Hopf curves in the $J_0,E$ plane, varying $J_1$ (C) and delay $D$ (D).
    \textbf{E-F)} Standing (E) and traveling (F) waves at $J_0=-2$, $J_1=-8$, $E=2$, $D=1$.
    \textbf{G)} sLIF simulation under same parameters as E-F.
    \textbf{H-I)} Turing-Hopf pattern beyond Turing and Hopf: $J_0=-60$, $J_1=20$, $E=10$, $D=0.2$.
    \textbf{J–S)} Additional spatiotemporal patterns at $E=10$, $D=0.2$:
    \textbf{J-K)} $J_0=-17$, $J_1=-60$;
    \textbf{L-M)} $J_0=-16$, $J_1=-72$;
    \textbf{N-O)} $J_0=-25$, $J_1=-95$;
    \textbf{P-Q)} $J_0=-5$, $J_1=-100$;
    \textbf{R-S)} $J_0=-15$, $J_1=-100$.
    }
    \label{fig:spatiotemporal}   
\end{figure}

Four primary bifurcations occur throughout the phase space when $D\neq0$ and $J_1\neq0$: the saddle-node, Hopf, Turing, and Turing-Hopf bifurcations (Fig.~\ref{fig:spatiotemporal}A,B: blue, green, red, purple respectively). 
The locations of these instabilities for any resting potential above threshold ($E>1$), are topologically similar to Fig.~\ref{fig:spatiotemporal}A with different scaling dependent on $E$ and the delay.
Similarly, the locations of the instabilities when $E<1$ are similar to Fig.~\ref{fig:spatiotemporal}B.

Formulaically, the Turing-Hopf curve \eqref{deltaTuringHopfEq} is similar to the Hopf curve \eqref{deltaHopfEq}.
This is because they are both instabilities with non-zero frequencies, but at different spatial modes.
Despite this, the Turing-Hopf curve increasingly resembles a reflection of the Turing curve across $J_1 = 0$ as the delay increases. This becomes exact in the limit $D \to \infty$.

Beyond the Turing-Hopf bifurcation, we observe the emergence of standing and traveling waves (Fig.~\ref{fig:spatiotemporal}E, F) induced by the $O(2)$ symmetry of $J$ to translations and reflections in $x$.
The sLIF exhibits both these patterns (Fig.~\ref{fig:delay}G). 
In that simulation, a standing wave spontaneously transitions into a traveling wave, suggesting metastability in the sLIF network.
Bistability of the standing and traveling waves with other states of the network remains to be investigated, and could be done using numerical continuation of solutions implementing both delays and spatial modulation.
This is the final region of the phase diagram we describe: \\

\begin{enumerate}
\item[x)] SW-TW: Standing and traveling waves exist and are stable. \\
\end{enumerate}

In regions beyond multiple primary instabilities, both the mean-field approximation and sLIF network exhibit a variety of additional spatiotemporal patterns, some likely arising from secondary or higher-order bifurcations.
Beyond both the Turing and Hopf bifurcations, we observe a Turing-Hopf pattern (Fig.~\ref{fig:spatiotemporal}H, I).
Beyond the Turing-Hopf bifurcation, we observe additional dynamics, likely due to an additional instability (Fig.~\ref{fig:spatiotemporal}P-Q).
Finally, beyond both the Turing-Hopf and Hopf bifurcations, we observe a variety of mixed mode oscillations (Fig.~\ref{fig:spatiotemporal}J-S).
Investigating these secondary instabilities and the rich variety of dynamical patterns to which they lead remains a direction for future research.

\begin{remark} The classic neural field equation $\partial_t x = -x + f(E + J \ast x)$, on the ring, has similar transitions from a homogeneous equilibrium to oscillatory and Turing patterns in the presence of delays and spatially modulated coupling~\cite{roxinRoleDelaysShaping2005}; see also~\cite{amariDynamicsPatternFormation1977, ben-yishaiTheoryOrientationTuning1995, wilsonExcitatoryInhibitoryInteractions1972, wilson_mathematical_1973}. 
The locations of those bifurcations differ in the classic rate and sLIF networks, however. For example, in the classic neural field equation the Turing instability of the homogenous active state occurs at $J_1 = 2$; in the sLIF mean-field dynamics we studied, the Turing instability occurs at $2 J_1 = J_0 + \sqrt{J_0^2 + 4(E-J_0)}$ \eqref{Turing}. Secondary bifurcations and codimension-2 points also differ between the two models.
\end{remark}

\section{Discussion}

Patterns in neural activity are classically modeled using neural field equations such as the Wilson-Cowan and Amari-Grossburg models \cite{ amariDynamicsPatternFormation1977, grossbergLearningEnergyentropyDependence1969, wilson_mathematical_1973}.
These can be derived as explicit mean-field theories for highly simplified microscopic models or by making strong assumptions like a separation of timescales between neural and synaptic dynamics 
\cite{buiceFieldtheoreticApproachFluctuation2007, ginzburgTheoryCorrelationsStochastic1994,
ohiraMasterequationApproachStochastic1993,
pintoQuantitativePopulationModel1996}. 
This complicates their relation to biophysical microscopic models. 
This discrepancy is one motivation for the development of next-generation neural field theories from specific microscopic models \cite{ byrneNextgenerationNeuralField2019, montbrioMacroscopicDescriptionNetworks2015, schwalgerMindLastSpike2019}. 

Soft-threshold integrate-and-fire networks replace the nonlinear dynamics of spike emission with a probabilistic spike-and-reset rule, but other biophysical detail can be directly incorporated. 
This family of models is often studied with a population density approach, which exposes rigorous mean-field limits \cite{ 
demasiHydrodynamicLimitInteracting2015,
gerstnerPopulationDynamicsSpiking2000, jabin_mean-field_2025}. 
The mean-field limit can also be exposed through the joint density functional of the network's state trajectories, rather than the population density \cite{ ockerDynamicsStochasticIntegratefire2023, robertDynamicsRandomNeuronal2016}. 
The population density approach has recently been extended to networks with either spatial connectivity or delays, exposing Hopf or Turing bifurcations separately \cite{cormierHopfBifurcationMeanField2021, dumontOscillationsFullyConnected2023,
dumontPatternFormationSpiking2024,
jabinDenseNetworksIntegrateandfire2024}.
% This family of models exposes rigorous mean-field limits 
% \cite{robertDynamicsRandomNeuronal2016}.

Here, we instead used a deterministic approximation to study the dynamics of soft-threshold leaky integrate-and-fire (sLIF) networks with both synaptic delays and spatial connectivity. This approximation has a similar form to the classic Amari-Grossberg equations with an additional term due to the voltage resets after action potentials.
We found oscillatory, spatial, and spatiotemporal instabilities that generate coherent activity patterns such as bulk oscillations, stationary Turing patterns, standing and traveling waves.
We identified various multi-stable regions of different network states due to the sub-criticality of the Hopf and Turing bifurcations.
In these regions, the network can support both patterned behavior and a homogeneous state (quiescent or active), and can be switched from one state to the other by global or local perturbations.
We confirmed all these predictions of the mean-field dynamics in simulations of the underlying microscopic stochastic system.

% Removed paragraph from here

From a mathematical point of view, all of the primary bifurcations we described could be studied using a parameter dependent infinite-dimensional center manifold reduction \cite{avitabile2020local, haragus2010local, vanderbauwhede1992center, veltz13}.  
Due to the lack of other positive spectrum of the homogeneous equilibrium at the bifurcation, invariant foliation results would give that stability on the manifold implies local stability in the full system.

The sLIF networks we studied are members of a broader family of soft-threshold integrate-and-fire networks with richer voltage dynamics. 
Soft-threshold integrate-and-fire networks with nonlinear voltage dynamics can have oscillatory solutions 
\cite{cormierHopfBifurcationMeanField2021}. 
Adaptive exponential integrate-and-fire neurons can capture a broad range of single-neuron spike patterns 
\cite{naudFiringPatternsAdaptive2008,touboulDynamicsBifurcationsAdaptive2008}. 
Adaptive nonlinear soft-threshold integrate-and-fire networks are also amenable to the same type of mean-field approximation we used here.
Including excitatory and inhibitory cell types and their relative spatial coupling profiles and synaptic timescales may also give rise to new spatiotemporal dynamics
\cite{huangCircuitModelsLowDimensional2019}.

Furthermore, it would be of interest to consider these patterns on unbounded domains as well as infinite dimensions. Near the Turing instabilities found here, we expect a family of (subcritical) periodic waves to bifurcate. 
The subcriticality and bistability in the bounded-domain problem indicate that it should be possible, in the unbounded domain, to construct fronts connecting the periodic state to the quiescent state, spatial patterns, and other complex modulated waves \cite{faye2015modulated, faye2018center}.

Next-generation neural field theories also aim to explicitly describe fluctuations, correlations, and synchrony \cite{byrneNextgenerationNeuralField2019}.
The neural field equations we study can be straightforwardly extended to include these through a fluctuation expansion of the density functional of the underlying microscopic model
\cite{ bressloffStochasticNeuralField2010, buiceFieldtheoreticApproachFluctuation2007, ockerDynamicsStochasticIntegratefire2023}. 
For the sLIF networks studied here, there are three possible types of fluctuation correction: 1) corrections to the mean-field rate $f(\bar{v})$ due to fluctuations in the voltage, 2) a correction to the mean-field voltage dynamics due to the spike-voltage covariance, and 3) variability in the synaptic field due to correlated or finite-size fluctuations in the activity.
These have each been studied in non-spatial networks
\cite{ockerDynamicsStochasticIntegratefire2023, paliwalMetastabilityNetworksStochastic2025, schmutzFiniteSizeNeuronalPopulation2023}.
How these different fluctuations interact with neural nonlinearities to shape spatiotemporal patterns in sLIF networks remains to be investigated.

\section*{Declarations}
\paragraph{Authorship:}
All authors have made substantial intellectual contributions to the study conception, execution, and design of the work. All authors have read and approved the final manuscript.  In addition, the following contributions occurred:  Conceptualization: Ryan Goh, Gabriel Koch Ocker; Formal analysis and investigation: Lauren Forbes, Jared Grossman, Ryan Goh; Writing - original draft preparation: Lauren Forbes; Writing - review and editing: Lauren Forbes, Ryan Goh, Gabriel Koch Ocker; Supervision: Montie Avery, Ryan Goh, Gabriel Koch Ocker.

\paragraph{Conflicts of interest:} 
The authors declare no conflicts of interest.

\paragraph{Data and code availability:}
Code to reproduce the results of this manuscript can be found at \href{https://github.com/lcforbes4/sLIFspatioTemporal}{https://github.com/lcforbes4/sLIFspatioTemporal}

\paragraph{Funding:}
The research of the authors was partially supported by NSF-DMS 2307650
(RG), and by a grant from the Allen Institute Mindscope Phase 4 program (GKO).

\newpage
\section{Appendix}

\subsection{Further reading}
We used largely standard methods for linear stability analysis of partial differential equations and numerical continuation, as described in \cite{coombesNeuralFieldsTheory2014, bressloffSpatiotemporalDynamicsContinuum2012, uecker2021numerical}. 
In the following sections, we describe these computations in detail.

\subsection{General dispersion relation}
\label{sec: General Dispersion Relationship}
To derive the dispersion relationship, we consider the mean-field approximation with a general coupling function $J(x,t-D)$ where $J(x,t)\in L^1((-\pi,\pi)\times(0,\infty))$ along with the intensity function $f(v(x,t))$,
\begin{equation}
    \partial_t v(x, t) = -v(x, t) + E - f(x, t) \, v(x, t)+\int_{-\infty}^{t-D} \int_{-\pi}^\pi J(x-y, t-s-D) \, f(y, s) \, dy \, ds \, 
    \label{general_mf}
\end{equation}
with periodic spatial domain, $x\in(-\pi,\pi]$.

Suppose the stationary uniform solution $v_0$, satisfying 
$v_0 = E - v_0f(v_0) + (J\ast f(v_0))(x,t-D)$, lies away from 
any non-differentiable points of $f$. Writing $v(x,t) = v_0 + w(x,t)$ 
with $w(x,t)$ sufficiently small, linearizing \eqref{general_mf} 
about $v_0$ yields
\begin{equation}
    \partial_t w = \left(-1  - f(v_0) - v_0f'(v_0)\right)w+ f'(v_0)\int_{-\infty}^{t-D}\int_{-\pi}^{\pi} J(x-y, t-s-D) w(y,s)\, dy\,ds
    \label{gen_linear}
\end{equation}
Taking the Fourier transform of \eqref{gen_linear} in space results in the following system of differential equations indexed by the wave number $k$
\begin{equation}
    \partial_t \hat{w}_k = (-1  - f(v_0) - v_0f'(v_0))\hat{w}_k+ f'(v_0)\int_{-\infty}^{t-D} \hat{J}_k(t-s-D) \hat{w}_k(s)\,ds\text{ , } k\in\mathbb{Z}
\end{equation}
where $\hat{w}_k(t) = \int e^{-ikx}w(x,t)\,dx$ and $\hat{J}_k(t)=\int e^{-ikx}J(x,t)\,dx$ are the time-dependent $k^{th}$ Fourier coefficients of $w$ and $J$. Finally, making the ansatz $\hat{w}_k=e^{\lambda_k t}$ and the substitution $\tilde{s}=t-s-D$, we get the dispersion relation
\begin{equation}
    \lambda_k = -1  - f(v_0) - v_0f'(v_0)+ f'(v_0)e^{-\lambda_k D}\int_{0}^{\infty} \hat{J}_k(\tilde{s}) e^{-\lambda_k \tilde{s}}\,d\tilde{s}
    \label{general dispersion}
\end{equation}
With the additional assumption of a threshold linear intensity function of the form $f(v)=\lfloor v-1\rfloor_+$, \eqref{general dispersion} simplifies to \eqref{general dispersion1}.

\subsection{Instabilities with delayed pulse coupling}
\label{sec:Instabilities with Delayed Pulse Coupling}

To derive the dispersion relation for delayed pulse coupling, we assume $\hat{J}_k(t)=\hat{J}_k\delta(t)$, $f(v)=[v-1]_+$, and $v_0>1$.  The dispersion relation \eqref{general dispersion} then simplifies to
\begin{equation}
    \lambda_k = -2v_0 + \hat{J}_ke^{-\lambda_k D}
    \label{delta_dispersion2}
\end{equation}
Separating the real and imaginary components of \eqref{delta_dispersion2} and assuming that $Re(\lambda_k) = 0$ yields
\begin{align}
    0 &= -2v_0 + \hat{J}_k \cos(-D\lambda_{i,k}) \label{delta_dispersion_real}
    \\
    \lambda_{i,k}  &= \hat{J}_k \sin(-D\lambda_{i,k}) \label{delta_dispersion_imag}
\end{align}
where $\lambda_{i,k}:=Im(\lambda_k)$.
The eigenvalue equation can also be formulated with the Lambert W function to achieve the same results \cite{corlessLambertWFunction1996}.
From now on, we assume the Fourier coefficients of cosine spatial coupling \eqref{cosinecoupling} which are $\hat{J}_0=J_0$, $\hat{J}_{\pm1}=J_1/2$, and $\hat{J}_{k}=0\,\forall k\notin\{0,\pm1\}$.
We next identify the locations of four instabilities of the stationary uniform equilibria $v_\pm$ given in \eqref{homog_solns}.

\vspace{0.5cm}
\noindent
\textit{I. Saddle-node Bifurcation ($k=0, \lambda_{i,0}=0$):} This is an instability of the uniform spatial mode with zero temporal frequency.
The imaginary part of the eigenvalue is zero, so \eqref{delta_dispersion_imag} is trivial. 
By substituting the form of $v_{\pm}$ given in \eqref{homog_solns}, \eqref{delta_dispersion_real} simplifies to $J_0=2\pm2\sqrt{1-E}$. 
If $J_0 = 2-2\sqrt{1-E}$, the values of $v_\pm$ are both below the threshold in $f$, an so neither equilibrium exists.
Therefore the instability only occurs along the curve
\begin{equation}
    J_0=2+2\sqrt{1-E}
    \label{delta_saddle}
\end{equation}

\vspace{0.5cm}
\noindent
\textit{II. Hopf Bifurcation ($k=0, \lambda_{i,0}\neq0$):} This is an instability of the uniform spatial mode with non-zero temporal frequency. 
We solve \eqref{delta_dispersion_real}, yielding $\lambda_i = \frac{-1}{D}\cos^{-1}(\frac{2v_0}{J_0})$, using the principle branch of $\cos^{-1}$. Substituting this into \eqref{delta_dispersion_imag} gives the Hopf curve as an implicit function of the parameters:
\begin{equation}
    \frac{-1}{D}\cos^{-1}\left(\frac{2v_0}{J_0}\right)  = J_0 \sqrt{1 - \left(\frac{2v_0}{J_0}\right)^2}.
    \label{Hopf_curve_appendix}
\end{equation}

We next show that only $v_+$ can undergo a Hopf Bifurcation and it requries $3J_0^2+4J_0 > 4E$ and $J_0<0$. 
Using \eqref{homog_solns}, the real part of the dispersion relation \eqref{delta_dispersion_real} simplifies to 

\begin{equation}
    1 \pm \frac{|J_0|}{J_0}\sqrt{1+\frac{4(E-J_0)}{J_0^2}}= \cos(-D \lambda_i)
    \label{real11}
\end{equation}

Due to the range of cosine, \eqref{real11} is only well defined for the minus case (instability of $v_-$) if $J_0>0$.
\eqref{real11} is well defined for the plus case (instability of $v_+$) if $J_0<0$.
But since the principle branch of arccosine is a non-negative function, the Hopf curve given in \eqref{Hopf_curve_appendix} is only well defined for $J_0<0$. 
Therefore, we conclude that only $v_+$ can undergo a Hopf bifurcation.

Finally, to satisfy the domain of arccosine, we have the condition
\begin{equation}
    \left|\frac{2v_0}{J_0}\right| = \left|\frac{J_0 + \sqrt{J_0^2 +4(E -J_0)}}{J_0}\right| < 1
\end{equation}
which can be rewritten as the set of inequalities $3J_0^2+4J_0 > 4E$, $v_{+}>1$, and $J_0<0$.

Using implicit differentiation, one can also show the transverse crossing condition of a Hopf bifurcation is satisfied along the curve \eqref{Hopf_curve_appendix}. 
As the delay is increased, the Hopf curve increases in $J_0$ and approaches the curve $3J_0^2+4J_0=4E$, which is the boundary of the region on which the Hopf curve \eqref{Hopf_curve_appendix} is well-defined.

\vspace{0.5cm}
\noindent
\textit{III. Turing Bifurcation ($k=1, \lambda_{i,1}=0$):} This is an instability of a non-uniform spatial mode with zero temporal frequency. Similar to the saddle-node bifurcation case, since the imaginary part of the eigenvalue is zero, \eqref{delta_dispersion_imag} is trivial and \eqref{delta_dispersion_real} simplifies to $2v_0 = J_1/2$. Note that $v_0>1$ only if $J_1>4$.
Therefore, there is a spatial instability of $v_\pm$ at
\begin{equation}
    \pm\sqrt{J_0^2+4(E-J_0)} = J_1/2 - J_0\text{ if }J_1>4
\end{equation}
This can be rewritten as
\begin{equation}
    E = J_1^2/4+J_0(4-J_1)
    \label{delta_turing}
\end{equation}
which corresponds to the Turing Bifurcation of $v_+$ if $J_0<J_1/2$ and a secondary instability of $v_-$ if $J_0>J_1/2$.

\vspace{0.5cm}
\noindent
\textit{IV. Turing-Hopf ($k=1, \lambda_{i,1}\neq0$):} This is an instability of a non-uniform spatial mode with non-zero temporal frequency. Similar to the Hopf bifurcation case, we substitute $\lambda_{i,1} = \frac{-1}{D}\cos^{-1}(\frac{2v_0}{J_0})$ from \eqref{delta_dispersion_real} into \eqref{delta_dispersion_imag}. This gives the curve of a spatiotemporal instability of $v_0$
\begin{equation}
    \label{delta_turing_hopf}
    \frac{-1}{D}\cos^{-1}\left(\frac{4v_0}{J_1}\right)  = \frac{J_1}{2} \sqrt{1 - \left(\frac{4v_0}{J_1}\right)^2}
\end{equation}
if $v_0>1$ as well, which occurs if $\frac{J_1}{4}\cos(D\lambda_{i,1})>1$ from \eqref{delta_dispersion_real}. 

Note that this curve intersects the saddle-node curve in a codimension-2 bifurcation if $J_0=\frac{J_1}{2}\cos(-D\lambda_{i,1})$. The instability \eqref{delta_turing_hopf} is a Turing-Hopf bifurcation of $v_+$ if $J_0<\frac{J_1}{2}\cos(-D\lambda_{i,1})$ and a secondary instability of $v_-$ if $J_0>\frac{J_1}{2}\cos(-D\lambda_{i,1})$.

\subsection{More realistic synaptic responses}
\label{sec: More Realistic Synaptic Temporal Responses}
\begin{figure}[h]
    \centering
    \includegraphics[width=\textwidth]{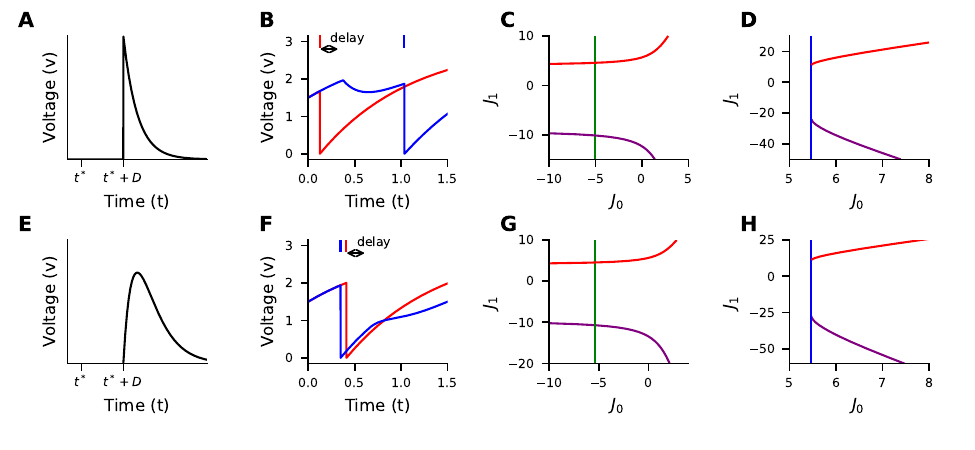}
    \caption{\label{fig:otherdelays}
    Instabilities with other temporal coupling profiles.
    \textbf{A)} The delayed exponential profile \eqref{delayedExponential}
    \textbf{B)} presynaptic neuron (red) emits a spike causing a change in membrane voltage of postsynaptic neuron (blue) after the delay.
    \textbf{C)} Primary instabilities present above threshold with the delayed exponential: Hopf(green), Turing(red), Turing-Hopf(purple), $E=2$, $D=1$, and $\tau=1$.
    \textbf{D)} Primary instabilities present below threshold with the delayed exponential: Saddle(blue), Turing(red), Turing-Hopf(purple), $E=-2$, $D=1$, and $\tau=1$. 
    \textbf{E)} The delayed alpha function (\ref{delayedAlphaFunction}).
    \textbf{F)} Presynaptic neuron (red) emits a spike causing a change in membrane voltage of postsynaptic neuron (blue) after the delay.
    \textbf{G)} Primary instabilities present above threshold with the alpha function: Hopf(green), Turing(red), Turing-Hopf(purple), $E=2$, $D=1$, and $\tau=1$.
    \textbf{H)} Primary instabilities present below threshold with the alpha function: Saddle(blue), Turing(red), Turing-Hopf(purple), $E=-2$, $D=1$, and $\tau=1$.
    }
\end{figure}

Synaptic potentials are characterized by three time scales: the latency or transmission delay (the time between the presynaptic spike emission and the postsynaptic response), as well as the rise time and the decay time \cite{destexheKineticModelsSynaptic1998}.
We consider two additional forms for the temporal coupling which model the rise and decay times of the synaptic potential in addition to the delay.
The first is the delayed exponential
\begin{equation}
    J_{\mathrm{time}}(t-D) = \frac{J_0}{\tau}e^{-(t-D)/\tau}H(t)
    \label{delayedExponential}
\end{equation}
where $H$ is the Heaviside step function.
This kernel implements a decay time of the action potential in addition to the delay (Fig.~\ref{fig:otherdelays}A).
The additional decay time smooths the postsynaptic response to incoming spikes compared to pulse coupling (Fig.~\ref{fig:otherdelays}B).
The locations of the four primary instabilities are very similar to the pulsed coupling case (Fig.~\ref{fig:otherdelays}C and D).
See Appendix \ref{sec: Instabilities with Delayed Exponential Coupling} for the derivation of the dispersion relations and instability curves.

The delayed alpha function (Fig.~\ref{fig:otherdelays}E) is another option for the temporal profile which also includes the rise time of the synaptic potential.
It is defined by
\begin{equation}
    J_{\mathrm{time}}(t-D) = \frac{J_0}{\tau^2}(t-D)e^{-(t-D)/\tau}H(t-D).
    \label{delayedAlphaFunction}
\end{equation}
This synaptic kernel further smooths the change in voltage of the postsynaptic neuron upon the arrival of a spike (Fig.~\ref{fig:otherdelays}F).
Again, the instability diagrams are qualitatively similar to the two previous cases of temporal kernels (Fig.~\ref{fig:otherdelays}G and H).
See Appendix \ref{Instabilities with Alpha Function Coupling} for the derivation of the dispersion relations and instability curve.
Notably, a nonzero delay is not required for the existence of a Hopf bifurcation and homogeneous oscillatory behavior with this temporal profile. 
The rise time of the alpha function acts as an effective delay in the transmission of action potentials between neurons.

%\todo{This is more speculative, but do we have more to say about the additional kernels? Just that we observe similar instability diagrams and hence similar temporal dynamics. Maybe mention (if true) that we expect (and observed?) qualitatively similar spatiotemporal dynamics when this temporal kernel was paired with the spatial coupling kernel used above.}

\subsubsection{Instabilities with delayed exponential coupling}
\label{sec: Instabilities with Delayed Exponential Coupling}
To identify the locations of instabilities with the delayed exponential temporal profile of the coupling, ie $\hat{J}_k(t)=\hat{J}_k \tau^{-1} e^{-(t)/\tau}H(t)$, we assume $f(v)=[v-1]_+$ and $v_0>1$. The dispersion relation \eqref{general dispersion} then simplifies to
\begin{equation}
    \lambda_k = -2v_0 + \hat{J}_k e^{-\lambda_k D}\frac{1}{(1+\lambda_k\tau)}
    \label{delayed_exp_dispersion}
\end{equation}
Separating the real and imaginary components of \eqref{delayed_exp_dispersion} with the assumption $Re(\lambda_k)=0$ results in

\begin{subnumcases}{}
    -\tau\lambda_{i,k}^2 = -2v_0 + \hat{J}_k\cos(\lambda_{i,k} D) \label{delayed_exp_dispersion_real}
    \\
    \lambda_{i,k} = \hat{J}_k\sin(-\lambda_{i,k}  D) -2v_0 \lambda_{i,k} \tau \label{delayed_exp_dispersion_imag}
\end{subnumcases}

Note that when $\lambda_{i,k}=0$, \eqref{delayed_exp_dispersion} is equivalent to \eqref{delta_dispersion2}. Therefore the instabilities with zero temporal frequency occur along the same curves as the delayed pulse case. The saddle-node bifurcation in the delayed exponential case is then also given by \eqref{delta_saddle} and the Turing bifurcation is given by \eqref{delta_turing}.

To identify the locations of instabilities with non-zero temporal frequencies, we first rewrite \eqref{delayed_exp_dispersion_imag} as $\lambda_i  = \hat{J}_k\sin(-\lambda_{i,k}  D)/(1+2v_0 \tau)$ and substitute it into \eqref{delayed_exp_dispersion_real} to get
$$-\tau\left(\frac{\hat{J}_k\sin(-\lambda_{i,k}  D)}{1+2v_0 \tau}\right)^2 = -2v_0 + \hat{J}_k\cos(\lambda_{i,k} D).$$
Then, by using the identity $\sin^2(x)=1-\cos^2(x)$ and defining $y:=\cos(\lambda_{i,k} D)$, we can transform the equation above into the quadratic
$$(-\hat{J}_k^2\tau)y^2 + (\hat{J}_k+4v_0\tau \hat{J}_k + 4v_0^2 \tau^2 \hat{J}_k)y + (\tau \hat{J}_k^2 -2v_0 -8 v_0^2\tau -8v_0^3\tau^2) = 0.$$
We solve this quadratic for $y$ and thus for $\lambda_{i,k}=D^{-1}\arccos(y),$
\begin{equation}
        \begin{aligned}
            &\lambda_{i,k} = \frac{1}{D}\arccos\Biggl(\frac{(1+4v_0\tau  + 4v_0^2 \tau^2 )}{2 \hat{J}_k\tau}
            \\
            &\pm\frac{\sqrt{(1+4v_0\tau + 4v_0^2 \tau^2 )^2+4\tau(\tau \hat{J}_k^2 -2v_0 -8 v_0^2\tau -8v_0^3\tau^2)}}{2 \hat{J}_k \tau}\Biggr).
        \end{aligned}
    \label{delayed_exp_lambda}
\end{equation}
Plugging this $\lambda_i$, which is in terms of the parameters $\hat{J}_k, E, \tau, D$, into the imaginary part of our dispersion relation \eqref{delayed_exp_dispersion_imag} gives
\begin{align}
        \arccos\left(\frac{A \hat{J}_k \pm |\hat{J}_k| \sqrt{A^2+4\tau(\tau \hat{J}_k ^2 -2v_0A)}}{2\hat{J}_k^2\tau}\right)(1+2v_0\tau)&\notag
        \\
        =-\hat{J}_k D\sqrt{1-\left(\frac{A\hat{J}_k\pm|\hat{J}_k|\sqrt{A^2+4\tau(\tau \hat{J}_k^2 -2v_0A)}}{2\hat{J}_k^2\tau}\right)^2}&,
    \label{delayed_exp_temporal_instab}
\end{align}
where $A=1+4v_0\tau + 4v_0^2 \tau^2$. 
This is the Hopf curve when $k=0$ and $\hat{J}_k=J_0$ or the Turing-Hopf curve when $k=1$ and $\hat{J}_k=J_1/2$.
    
Next, we will look at which parameter values and which equilibria $v_0$ the curve \eqref{delayed_exp_temporal_instab} is well defined for. 
First, we note that the range of the principle branch of arccosine is non-negative, so \eqref{delayed_exp_temporal_instab} is not defined for $\hat{J}_k>0$.

To satisfy the domain of arccosine in \eqref{delayed_exp_lambda} to be defined, we need
$$\left|\frac{A\pm\sqrt{A^2+4\tau(\tau \hat{J}_k^2 -2v_0A)}}{2\hat{J}_k\tau}\right|\leq1.$$ 
When $k=0$, the minus case does not satisfy this inequality and the plus case can be rewritten as the set of conditions. $J_0<-2$, $v_+>1$, and $4E<4J_0+3J_0^2$.
Therefore the Hopf and Turing-Hopf bifurcations are both instabilities of $v_+$.

\subsubsection{Instabilities with alpha function coupling}
\label{Instabilities with Alpha Function Coupling}
To identify the dispersion relation and primary instabilities of the homogeneous solution with the delayed alpha function chosen for the temporal coupling profile, we assume $v_0>1$, $f(v)=[v-1]_+$ and $\hat{J}_k(t)=\frac{\hat{J}_k}{\tau^2}te^{-t/\tau}H(t)$. The dispersion relation \eqref{general dispersion} becomes 
\begin{equation}
    \lambda_k(1+\lambda_k\tau)^2 + 2v_0(1+\lambda_k\tau)^2 = \hat{J}_k e^{-\lambda_k D}
    \label{alpha_function_dispersion_relation}
\end{equation}
Separating the real and imaginary components of \eqref{alpha_function_dispersion_relation} with the assumption $Re(\lambda_k)=0$, 
\begin{align}
    &\lambda_{i,k}^2(- 2\tau - 2v_0\tau^2) = \hat{J}_k\cos(D\lambda_{i,k}) - 2v_0 \label{alpha_function_dispersion_relation_real}
    \\
    &\lambda_{i,k}(1 + 4v_0\tau - \lambda_{i,k}^2\tau^2) = -\hat{J}_k\sin(D\lambda_{i,k}) .\label{alpha_function_dispersion_relation_imag}
\end{align}
When $\lambda_{i,k} = 0$, as in the precious cases, the saddle-node and Turing bifurcations are given by \eqref{delta_saddle} and \eqref{delta_turing} respectively. 

Next we identify the other two types of instabilities which have non-zero temporal frequency. 
From the real part of the dispersion relation \eqref{alpha_function_dispersion_relation_real},
$$\lambda_i = \pm\sqrt{\frac{\hat{J}_k\cos(\lambda_i D)-2v_0}{-2\tau-2v_0 \tau^2}},$$
and substitute it into \eqref{alpha_function_dispersion_relation_imag},
\begin{equation}
    \left(\frac{\hat{J}_k\cos(\lambda_{i,k} D)-2v_0}{-2\tau-2v_0 \tau^2}\right)\left(1 + 4v_0\tau -  \left(\frac{\hat{J}_k\cos(\lambda_{i,k} D)-2v_0}{-2\tau-2v_0 \tau^2}\right)\tau^2\right)^2 = \hat{J}_k^2\sin^2(\lambda_{i,k} D).
\end{equation}
Then, using the identity $\sin^2x=1-\cos^2x$ and defining $y:=\cos(\lambda_{i,k} D)$, we have the equation
\begin{equation}
    \left(\frac{\hat{J}_ky-2v_0}{-2\tau-2v_0 \tau^2}\right)\left(1 + 4v_0\tau -  \left(\frac{\hat{J}_ky-2v_0}{-2\tau-2v_0 \tau^2}\right)\tau^2\right)^2 = \hat{J}_k^2(1-y^2),
\end{equation}
which can be rewritten as cubic in $y$     
\begin{multline}
    -\frac{\hat{J}_k^3\tau}{8(\tau v+1)^3}y^3+\frac{\hat{J}_k^2\left(4\tau^3v^3+4\tau^2v^2+5\tau v+2\right)}{4(\tau v+1)^3}y^2
    \\
    - \frac{\hat{J}_k(2\tau v+1)^2\left(4\tau^2v^2+2\tau v+1\right)}{2\tau(\tau v+1)^3}y+\frac{v(2\tau v+1)^4-\hat{J}_k^2\tau(\tau v+1)^3}{\tau(\tau v+1)^3}= 0.
\end{multline}
We can explicitly solve for the roots of the cubic, one of which is always real.

Finally, substituting $\lambda_{i,k}= \frac{1}{ D}\arccos(y)$ into the real equation \eqref{alpha_function_dispersion_relation_real} gives an implicit equation of the instability in terms of the system parameters,
\begin{equation}
    \left(\frac{1}{ D}\arccos(y)\right)^2(- 2\tau - 2v_0\tau^2) = \hat{J}_ky - 2v_0.
    \label{alpha_func_0_freq_instability_curve}
\end{equation}
This is the Hopf curve when $k=0$ and $\hat{J}_k=J_0$ and a Turing-Hopf curve when $k=1$ and $\hat{J}_k=J_1/2$.

\subsection{Numerical methods}
\label{sec: Numerical Methods}

\subsubsection{Simulations}

Simulations of the mean-field approximation \eqref{mean-field} were done using the forward Euler method with a step size of $dt=0.001$.
The ring $(0,2\pi]$ was descritized as $x_i=-\pi+\frac{2\pi}{N}i$, $i=1,...,N$, and $N=100$ ($N=1000$ for MF simulations of spatiotemporal dynamics in Fig.~\ref{fig:spatiotemporal}).

Simulations of the sLIF network \eqref{discreteLIF} were done using the stochastic forward Euler method with step size $dt=0.001$ with $N=1000$. 
The connectivity matrix $\mathbf{J}\in\mathbb{R}^{N\times N}$ was randomly sparse, with connection probability $p=0.5$. 
The connections between neuron pairs $(i,j)$ were independently sampled as Bernoulli random variables with probability $p=0.5$.
Each non-zero connection then has synaptic weight given by $J_{ij}=\frac{2\pi}{pN}(J_0+J_1\cos(x_i-x_j))$.
Self-connections are excluded by setting $J_{ii}=0$.
We generated spikes by sampling $dn_i$ as Bernoulli random variables at each time step with success probabilities $f(v_i) dt$.
Finally, synaptic delays were incorporated with a time-shift of $N_{\text{delay}}=\lfloor D/dt \rfloor$ to the synaptic input.

The initial conditions must be defined for $t\in[-D,0]$ due to the presence of the delay.
Most patterns were initialized with constant initial conditions or the stationary form $A+B\cos(x)$.
For the traveling wave pattern seen in Fig.~\ref{fig:spatiotemporal}F, we used an initial condition of the form $A+B\cos(x+tv)$ with $v\neq0$ to induce wave propagation.
Details for specific initial conditions are found in the code repository \href{https://github.com/lcforbes4/sLIFspatioTemporal}{https://github.com/lcforbes4/sLIFspatioTemporal}.

\subsubsection{Inducing network transitions in bistable regions}

To switch from the high activity state to the oscillatory state (Fig.~\ref{fig:deltaStochVsMean_bistab} I-J), a positive pulse with short duration was applied to the entire network.
To turn off the oscillatory state and return to the high activity state, a negative pulse was applied during the peak of the oscillations to suppress oscillatory activity.
The timing, amplitude, and duration of the negative pulse were fine tuned to the particular simulation and set of parameters to achieve the return to the high activity state.
See figure caption for specific pulse parameters.

In all three simulations (Fig.~\ref{fig:spatial-bistab}A–C), a spatially dependent drive of the form 
$-2\cos(x)$, applied for a duration of 1, was used to induce a transition to spatially patterned network activity.
To transition from the spatially patterned activity state to the high activity state (Fig.~\ref{fig:spatial-bistab}, A and C), a pulse with amplitude -2 was applied to the quiescent middle third of the network (neurons $N/3:2N/3$).
To transition from the spatially patterned activity state to the quiescent state (Fig \ref{fig:spatial-bistab} B), a pulse with amplitude -2 was applied to the active outer two thirds of the network (neurons $0:N/3$ and $2N/3:N$).

\subsubsection{Continuation of oscillatory solutions}

Continuation of oscillatory solutions was performed using a one-dimensional, spatially homogeneous mean-field reduction \eqref{mean-field}. 
This approximation assumes zero spatial modulation ($J_1=0$) in the coupling function, such that each neuron in the network evolves according to the spatially independent delay-ODE.
We used the \textsc{Matlab} package DDE-Biftool, which was developed for bifurcation analysis of delay differential equations (\cite{engelborghsNumericalBifurcationAnalysis2002,sieberDDEBIFTOOLManualBifurcation2016}).
First, we computed and continued a branch of equilibria along with the spectrum of its linearization to identify the location of the Hopf bifurcation.
Then we initialized a small-amplitude orbit near the Hopf bifurcation to construct and continue a branch of oscillatory solutions.
Stability of periodic orbits was determined by simultaneously computing Floquet multipliers of the linearized periodic map.
We found the Hopf bifurcation to be subcritical, with a branch of unstable oscillations emerging backwards out of the bifurcation point as $J_0$ and $E$ decrease and $D$ increases.

\subsubsection{Continuation of Turing patterns}

The periodic Turing patterns computed in Section~\ref{sec: Spatial Instabilities - Turing Bifurcation} are steady-state solutions of the mean-field approximation \eqref{mean-field} with spatial connectivity ($J_1\neq 0$) but without delay ($D=0$). 
To compute these solutions numerically in \textsc{Matlab}, we implemented a Newton-GMRES FFT method \cite{rankinContinuationLocalizedCoherent2014}.
We discretized \eqref{mean-field} using a Fourier spectral method with \(N=2^{12}\) modes on the periodic domain $[0,2\pi)$.
We denote the discretized time-independent version of \eqref{mean-field}, as the non-linear system $F_1(v)=0$ where $v\in\mathbb{R}^N$ is the spatially discretized steady-state solution.

To obtain a good initial guess for the Turing pattern, we ran a simulation initialized with a small perturbation (of order \(10^{-1}\)) of the spatially homogeneous steady state. 
To fix the continuous translational symmetry on the ring, we appended a phase condition on our system,
\[
F_2(v)=\left\langle u(x), \sin(x) \right\rangle_{L^2} = 0,
\]
which pins the solution, and modified $F_1$ with an additional drift term $c\partial_xv$. 
The imposition of this condition requires an additional unknown or dummy variable $c$.
We then solved the resulting augmented system
$F(v,c)=\begin{bmatrix}
    F_1(v) + c\partial_x v\\F_2(v)
\end{bmatrix}=0$
for $v$ and $c$ using a Newton–GMRES method, where vector products in the Jacobian were computed spectrally and GMRES without preconditioning was used to compute the Newton step.

Once the initial Turing pattern was obtained, we performed numerical continuation in a single bifurcation parameter $\beta$ using secant continuation.
To do this, we introduced the augmented variable $z=[v,c,\beta]$ and appended the orthogonality condition
\[
F_3(z)=\left\langle z - z_0, sec \right\rangle = 0,
\]
where \(z_0=[v_0,c_0,\beta_0]\) is the previously computed solution in the continuation and $sec\in\mathbb{R}^{N+2}$ is the normalized secant vector between the two most recent solutions.
The variable $c$ stays uniformly close to zero throughout the continuation. 
The complete augmented system used for secant continuation is then
$$F(z)=\begin{bmatrix}
    F_1(v)+ c\partial_x v\\F_2(v)\\F_3(z)
\end{bmatrix}=0\text{ where }z=[v,c,\beta].$$

\begin{figure}[h]
    \centering
    \includegraphics[width=0.6\textwidth]{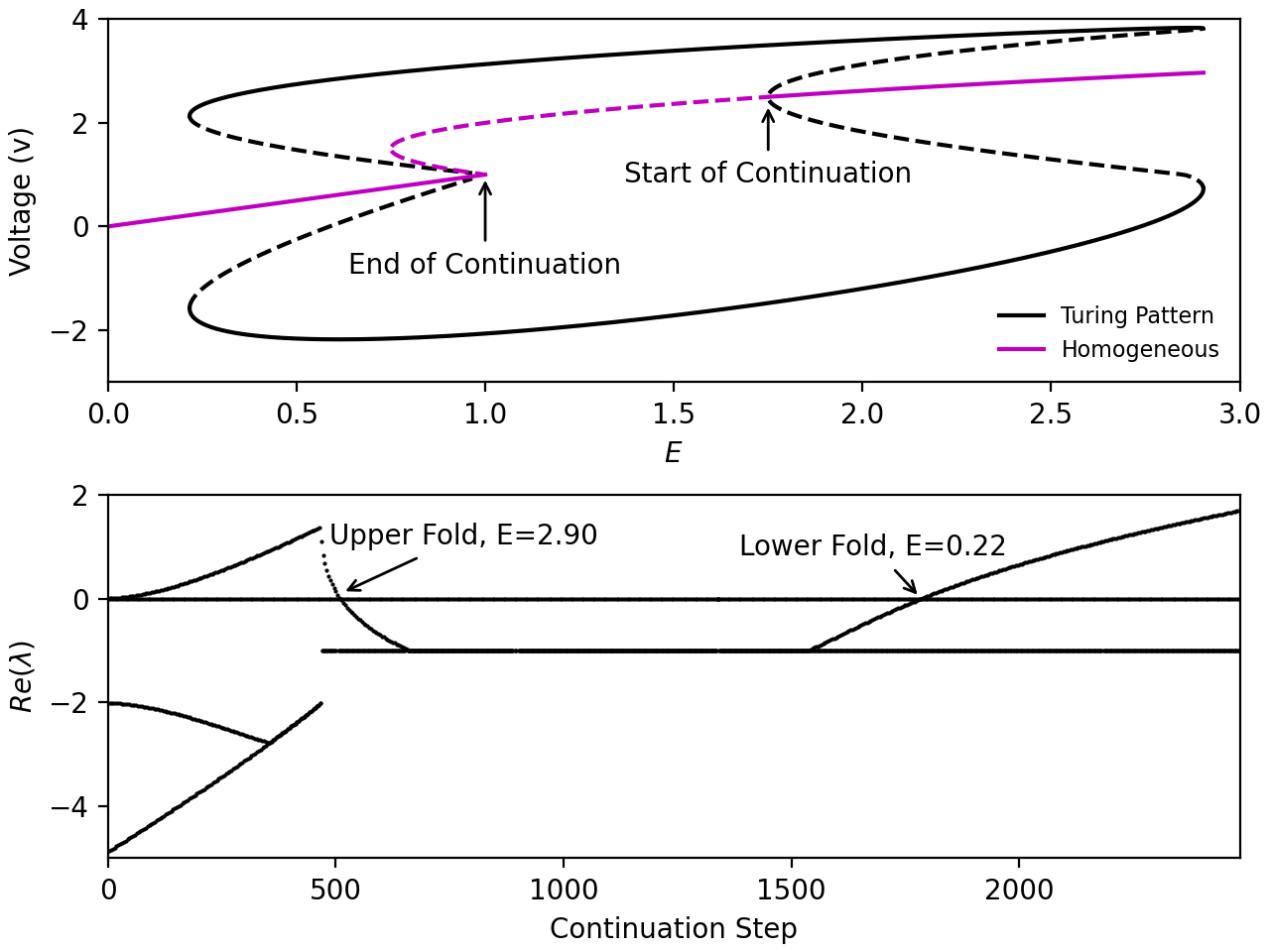}
    \caption{\label{fig:continutation_evals}
    Tracking eigenvalues along the continuation of the Turing pattern.
    }
\end{figure}

At each step of the continuation, we computed the spectral stability of the linearized operator near the origin with leading eigenvalues of Jacobian of the original system $F_1$ (without the phase and orthogonality conditions) and Matlab's eigs function. 
In Fig.~\ref{fig:continutation_evals}, we plot the real part of the eigenvalues with $Re(\lambda)>-5$ along the entire continuation curve in the parameter $E$.
The continuation starts at the Turing bifurcation and initially has an eigenvalue with $Re(\lambda)>0$, indicating instability and likely a subcritical bifurcation.
When the largest eigenvalue drops below zero (which occurs near $E=2.90$), this marks the location of a fold of Turing patterns and the continuation's switch onto the stable branch at the upper fold. 
The continuation then follows the stable branch until the lower fold where a real part of an eigenvalue becomes positive ($E=0.22$) and solutions are unstable until terminating at the non-smooth Turing bifurcation.
The eigenvalue with $Re(\lambda)=0$, which remains present throughout the continuation, is a consequence of the translational invariance of solutions on the ring.
The locations of the two fold curves in the phase diagram in Fig.~\ref{fig:spatial-bistab}D were identified by determining the location of the fold in multiple single-parameter continuations of the Turing pattern at varying parameters, not through continuation of the fold itself.

\subsection{Tracking the fold points to identify the boundaries of multi-stable regions}
\label{sec: Tracking the fold points to identify the boundaries of bistable regions}

\begin{figure}[h!]
    \centering
    \includegraphics[width=\textwidth]{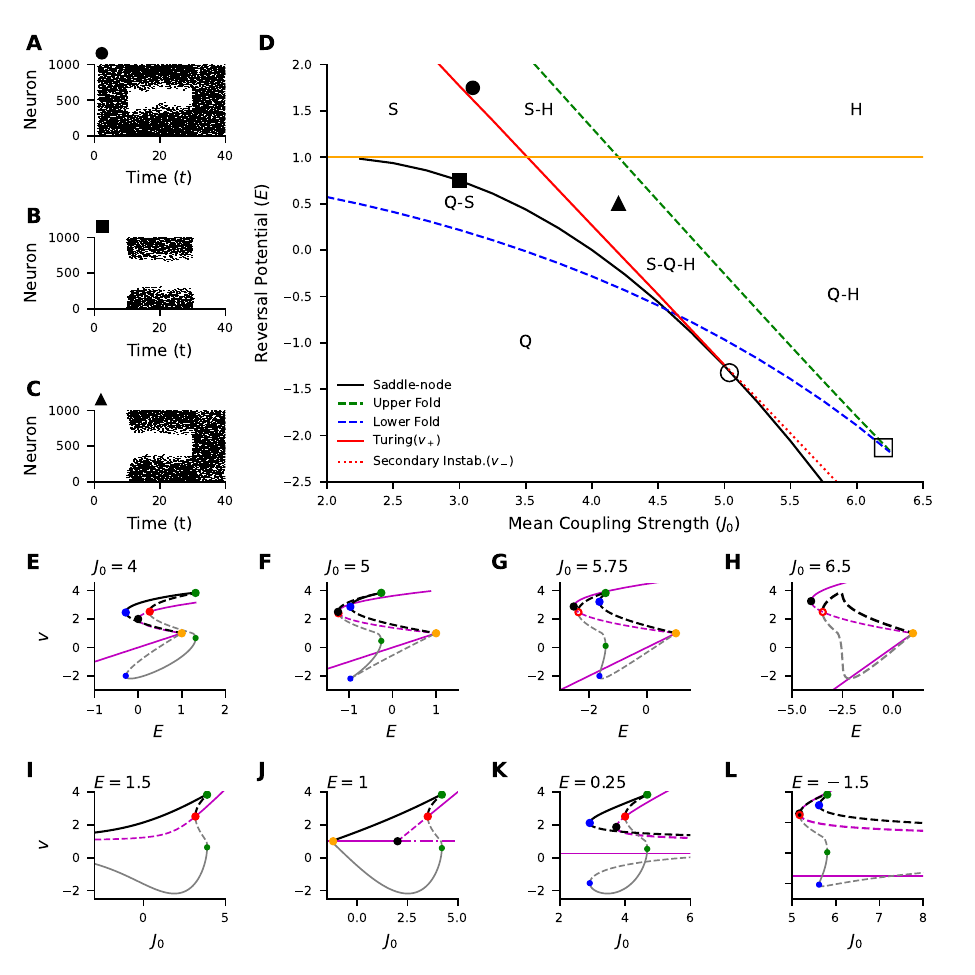}
    \caption{Regions of multi-stability near codimension-2 bifurcations with $J_1=10$.
    \textbf{A-C)} Raster plots showing multi-stability of spiking network's activity at parameter locations marked in panel D. At $t=15$ and $t=30$, perturbations were applied to $E$.
    \textbf{D)} Phase diagram in $J_0,E$ plane showing regions quiescent (Q), High (H), spatially patterned (S) activity, and  regions of bistability notated with a hyphens (e.g. Q-H) separated by various bifurcation curves. Turing bifurcation of $v_+$ (red, solid), secondary spatial instability of $v_-$(red, dashed), fold of Turing patterns (green and blue, dashed), saddle-node bifurcation (black, solid), and `non-smooth' saddle-node/Turing bifurcation (orange). The codimension 2 bifurcation of the simultaneous Turing and saddle-node bifurcations marked with an empty circle and the cusp-point of the two folds of Turing patterns marked with an empty square.
    \textbf{E-H)} Bifurcation diagrams in parameter $E$ at the $J_0$ value written at top of panel. Colored circles corresponding to the various bifurcations curves shown in the phase diagram.
     \textbf{I-L)} Bifurcation diagrams in parameter $J_0$ at the $E$ value written at top of panel.
    }
    \label{fig:spatial-bistab}   
\end{figure}

%The local activity regions (including regions of multi-stability with other states) are shown in the phase diagram Fig.~\ref{fig:spatial-bistab}D.
The regions Q-S and S-Q-H only exist in a small neighborhood near the co-dimension 2 bifurcation of the saddle-node and Turing bifurcations (Fig.~\ref{fig:spatial-bistab}D).
The boundaries of these and the surrounding regions are defined by the various instabilities which occur in this neighborhood: the subcritical Turing bifurcation (Fig.~\ref{fig:spatial-bistab}D, solid red) and nearby fold of Turing patterns (Fig.~\ref{fig:spatial-bistab}D, dashed green), the saddle-node bifurcation (Fig.~\ref{fig:spatial-bistab}D, solid black), and the subcritical non-smooth Turing (Fig.~\ref{fig:spatial-bistab}D, solid orange) with a nearby fold of the Turing patterns (Fig.~\ref{fig:spatial-bistab}D, dashed blue).
We refer to this second fold as the `lower' fold since it occurs at lower $E$ and $J_0$ values than the `upper' fold.
Since both of the spatial instabilities are subcritical in this neighborhood, the boundaries of the stable Turing pattern region are determined by the folds.

To track the fold points and understand the interaction of various instabilities in this region, we computed bifurcation diagrams across a range of parameter values (Fig.~\ref{fig:spatial-bistab}E-L).
These bifurcation diagrams are slices of the phase diagram (Fig.~\ref{fig:spatial-bistab}D) with an additional dimension showing $v$. 
The upper row (Fig.~\ref{fig:spatial-bistab}E-H), which shows continuation of the Turing pattern in the parameter $E$, are vertical slices of Fig.~\ref{fig:spatial-bistab}A at different $J_0$ values.
The lower row (Fig.~\ref{fig:spatial-bistab}I-L), showing continuation in $J_0$, are horizontal slices of Fig.~\ref{fig:spatial-bistab}A at various $E$ values.

In each of these bifurcation diagrams, the instabilities' locations are marked with a circle whose color matches the corresponding curve in the phase diagram (Fig.~\ref{fig:spatial-bistab}D).
For example, consider Fig.~\ref{fig:spatial-bistab}E, where five different instabilities (uniquely colored circles) are visible.
The saddle-node bifurcation, where $v_+$ and $v_-$ collide and vanish, is marked with a black circle.
The Turing bifurcation, where a unstable Turing pattern emerges from $v_+$, is shown with a red circle.
Both a non-smooth Turing and saddle-node bifurcation occur at the same point, shown with an orange circle, where $v_-$, $v_Q$, as well as an unstable Turing pattern emerge as $E$ is decreased.
Lastly, there are two folds of the Turing pattern:
the upper fold (green circle), connected to the Turing bifurcation via an unstable branch of Turing patterns, and the lower fold (blue circle), connected to the non-smooth Turing via an unstable branch.
The stable branch connecting the two folds determines where there is stable spatially patterned activity.

All three regions of multi-stability with the Turing pattern are present when $J_0=4$ (Fig.~\ref{fig:spatial-bistab}E).
First, the S-H region, where only the Turing pattern and $v_+$ are stable, lies between the non-smooth Turing bifurcation and the upper fold.
The Q-S region, where there is bistability of $v_Q$ and the Turing pattern, lies between the lower fold and the Turing bifurcation.
The last region of multi-stability between three solutions (the Turing pattern, $v_Q$, and $v_+$) lies between the Turing and non-smooth Turing bifurcations.

As $J_0$ increases (Fig.~\ref{fig:spatial-bistab}F, $J_0=5$), the S-H and Q-S regions vanish and S-Q-H is the only remaining multi-stable region with the Turing pattern.
The two fold points (green and blue) have moved closer together and passed by the locations of the spatial instability (red) and the non-smooth Turing (orange).
Therefore the stable branch of Turing patterns between them is smaller and lies entirely in the region where both $v_+$ and $v_Q$ are also stable.

In addition to the loss of two of the bistable regions at $J_0=4$, the location of the spatial instability is now on the $v_-$ branch after having just passed through the location of the saddle-node bifurcation (black circle) in a co-dimension 2 bifurcation. 
Despite being a secondary instability, the emerging branch of unstable Turing patterns leads to a stable branch through a fold (Fig.~\ref{fig:spatial-bistab}F, green). 
We mark this secondary instability of $v_-$ with an open red circle (largely obscured by the black circle) to distinguish it from the Turing bifurcation of $v_+$ (marked with a solid red circle in the other panels).

If $J_0$ is further increased (Fig.~\ref{fig:spatial-bistab}G, $J_0=5.75$), the distance between the two folds continues to decrease and thus the size of the S-Q-H region decreases.
They continue to approach each other until they collide and vanish in a codimension-2 cusp point (Fig.~\ref{fig:spatial-bistab}D, empty square) along with the stable branch between them, ending the S-Q-H region.
Beyond the cusp point, an unstable branch of Turing patterns remains, connecting the spatial instability to the non-smooth bifurcation (Fig.~\ref{fig:spatial-bistab}H, $J_0=6.5$).

We can also see the emergence and disappearance of these bistable regions as the resting potential passes through threshold by looking at the continuation of the Turing pattern in the parameter $J_0$, at differing $E$ values.
When above threshold ($E>1$, Fig.~\ref{fig:spatial-bistab}I), there is no lower fold and thus the only bistable region with spatially patterned activity is S-H.
The stable branch of Turing patterns exists for all $J_0$ values below the upper fold extending off to $-\infty$ and slowly decreasing in amplitude as $J_0$ is decreased.
The stable branch also decreases in amplitude as $E$ is decreased until at the threshold ($E=1$, Fig.~\ref{fig:spatial-bistab}J), 
it no longer extends to $-\infty$ and instead meets $v_Q$ at the non-smooth supercritical Turing bifurcation (Fig.~\ref{fig:spatial-bistab}J, $J_0\approx-1.3$).
This is where the non-smooth Turing bifurcation becomes subcritical in the bifurcation parameter $E$ and the lower fold emerges.
Recall there are multiple transitions between homogeneous solutions at threshold.
For $J_0<2$, the above threshold state $v_+$ meets the below threshold state $v_Q$ (i.e. $v_+=v_Q=1$).
For $J_0>2$, there is the non-smooth saddle bifurcation where both $v_-$ and $v_Q$ emerge.
Thus the homogeneous solution at $E=1$ is stable for $J_0<2$ and half-stable (dash-dotted) for $J_0>2$.
Then below threshold ($E<1$, Fig.~\ref{fig:spatial-bistab}K), there is a lower fold and an unstable branch of Turing patterns that extends off to $+\infty$ in $J_0$.
Consequently, there are now Q-S and S-Q-H regions.
As $E$ decreases, the two folds move closer (Fig.~\ref{fig:spatial-bistab}L) leaving only the S-Q-H region.
The folds eventually meet in the codimension-2 cusp point (Fig.~\ref{fig:spatial-bistab}A, blue dashed, green dashed, empty square) ending the branch of stable Turing patterns and spatially patterned activity region.

\bibliographystyle{siamplain}
\bibliography{references}
\end{document}